\documentclass[preprintnumbers, twocolumn,showkeys,superscriptaddress,showpacs,
nofootinbib,prd,notitlepage]
{revtex4-1}
\usepackage{bbm,pifont}
\usepackage[colorlinks=true, linkcolor=blue, citecolor=blue, urlcolor=blue]{hyperref}
\usepackage{amssymb}
\usepackage{amsmath}
\usepackage{physics}
\usepackage{enumitem}
\usepackage{amsfonts}
\usepackage{epsfig}
\usepackage{graphicx}
\usepackage{tabularx}
\usepackage{color}
\usepackage{slashed}

\begin{document}
\title{Emergence of non-Markovian Decoherent Histories in Integrable Environment: \\ A ``Tape Recorder'' Model for Local Quantum Observables}
\author{Nataliya Arefyeva}
\email[]{arefnat8@gmail.com}
\affiliation{Russian Quantum Center, 30 Bolshoy Boulevard, building 1, Skolkovo Innovation Center territory, Moscow, 121205, Russia} 
\affiliation{Physical Department, Lomonosov Moscow State University, Vorobiovy Gory, Moscow 119991, Russia} 
\author{Evgeny Polyakov}
\email[]{evgenii.poliakoff@gmail.com}
\affiliation{Russian Quantum Center, 30 Bolshoy Boulevard, building 1, Skolkovo Innovation Center territory, Moscow, 121205, Russia}
\begin{abstract}
We propose a new approach to coarse-grained description of quantum evolution that provides an explicit recipe to construct and evaluate multi-time decoherent histories in a controlled way, applicable to non-Markovian and integrable systems. Specifically, we study local interaction quench of a local degree of freedom (an open quantum system) within a noninteracting integrable environment. This setting allows us to identify the environmental degrees of freedom that irreversibly store records of the system's past. These modes emerge sequentially in time and define the projectors required for decoherent histories. We show numerically that the off-diagonal elements of the decoherence functional are exponentially suppressed relative to a significance threshold.
\end{abstract}

\maketitle

\section{Introduction}
The formalism of decoherent histories, also known as consistent histories and proposed in \cite{Griffiths:84, GMH:93, Omnes:87} provides a framework to understand the emergence of classical reality from quantum mechanics \cite{Halliwell:95}. A history is a sequence of projectors at different times, representing a coarse-grained description of the system’s dynamics \cite{GMH:QCG07, Griffiths:84}. For such histories to obey classical probability rules --- a property known as consistency --- the interference between distinct histories must be negligible. This is quantified by the decoherence functional \cite{DowkerHalliwell:92}, which must be approximately diagonal \cite{DowkerHalliwell:92, DowkerKent:96}. The approach has garnered significant interest, both for foundational questions in quantum theory \cite{Omnes:92, DowkerKent:96, Griffiths:24, Hohenberg:12, Griffiths:93} and for exploring the quantum-to-classical transition, including discussions about whether quantum dynamics can be reduced to a classical stochastic process \cite{Schmidtke:16, GemmStein:14, Szankowski:23, RieZur:16, Strasberg:class23, StDiaz:19, Lonigro:24, Milz:20, Smirne:18}.

Despite its conceptual appeal, the theory faces two primary challenges. First, the explicit construction of histories with a controlled degree of decoherence remains a computationally difficult problem \cite{Halliwell:99, Halliwell:93, Strasberg:24}. Second, the physical mechanism responsible for the emergence of decoherent histories is not fully understood \cite{Wang:25, Zurek:03, Ferte:25}.

Existing models fall into two broad categories. The first involves artificial ancilla (record) models \cite{Brun:00, Halliwell:99}, where a local system interacts sequentially with a series of abstract, non-physical subsystems. While useful, these models lack a connection to a physical environment. The second category comprises microscopic models with explicit Hamiltonians. Successful constructions in this category have largely been confined to Markovian open quantum systems \cite{PazZurek:93, Diosi:95, Brun:97, Brun:95, Brun:96} or to the quantum Brownian motion model \cite{Caldiera:83} via classical ensemble averaging \cite{BrunHartle:99, Halliwell:01, DowkerHalliwell:92, Subası:12}. For closed systems, numerical evaluations of histories exist \cite{BrunHall:96, Strasberg:24, Wang:25}, but computational complexity typically restricts them to a few time steps. This challenge has prompted studies on implementing decoherent histories on quantum computers \cite{Arrasmith:19}.

Histories satisfying the decoherence condition can be defined by various sets of projectors, corresponding to different coarse-grainings. The standard way is to define histories using coarse-grained projections onto subspaces associated with slow, global observables, analogous to how coarse-grainings are introduced in equilibrium statistical mechanics \cite{Halliwell:93, Halliwell:01, GMH:QCG07}. Recent numerical studies employing this method \cite{Strasberg:24, Wang:25} emphasize that non-integrability is a key factor for the emergence of decoherent histories, as it leads to a statistical suppression of the off-diagonal elements of the decoherence functional, resulting from the complex interference between microstates during the evolution. In contrast, these studies show a lack of convincing decoherence in integrable systems.

In this work, we propose an alternative mechanism that operates effectively in non-Markovian and integrable systems. Complementing the work of \cite{PolAref:24}, we shift the focus from global, slow observables to a local system interacting with an integrable environment. Our central principle is to treat the environment explicitly as a device that records the system's history. For a history to be decoherent, information about the local system must be stably and irreversibly stored in environmental degrees of freedom.

This idea resonates with the strong decoherence condition of Gell-Mann and Hartle \cite{GMH:98}, which requires that coarse-grained records of the past be preserved. The physical picture is that different event branches leave distinct, orthogonal traces in the environment (e.g., photons escaping to infinity), whose mutual orthogonality suppresses interference. However, in a spatial setting, this decoupling often proceeds slowly (e.g., as a power law), hindering practical implementation.

Our key insight is that by reformulating the problem in a time-dependent basis, we observe exponentially faster temporal decoupling. This allows us to realize the Gell-Mann and Hartle picture in a concrete dynamical model. We introduce an analogy with a magnetic tape recorder: the local observable acts as a read-write head, while the environment constitutes the tape. In a simple tape recorder model the head's field is long-range and decays with a power law, meaning it formally interacts with the entire tape at once. The crucial step is to introduce a finite sensitivity (or significance) threshold, analogous to a recorder's inability to read a signal below its noise floor. This threshold effectively localizes the interaction, yielding a process where information is sequentially and irreversibly recorded on distinct, passing segments of the tape. In our physical setting, this translates to a dynamic classification of the environmental modes into the following groups: (1) modes currently interacting with the system (under the ``head''), (2) modes that have irreversibly decoupled, now storing stable records of the past on the ``tape'', (3) modes that never significantly interact, and (4) modes which have not yet interacted but are scheduled to do so in the future (the segment of tape that will later pass under the head). It is the irreversibly decoupled modes --- the written segments of the tape --- that provide the stable, orthogonal records which define the projections for decoherent histories.

This ``tape recorder'' mechanism extends concepts from quantum Darwinism \cite{RieZur:16, Ollivier:04, Zurek:09}. While quantum Darwinism focuses on the redundant encoding of instantaneous states in environmental fragments, our mechanism encodes multi-time histories via a sequence of temporally decoupled modes.

We demonstrate this by providing an explicit algorithm for constructing histories with controlled consistency. The decoherence improves exponentially as the significance threshold is lowered, analogous to tuning the effective range of the recording head field in the tape recorder analogy.

The paper is structured as follows. Section~\ref{sec:Dec_His} provides preliminaries on decoherent histories and introduces the tape recorder model. Section~\ref{sec:model} describes the physical system and its coarse-grained description. In Section~\ref{sec:comp_fram}, we introduce a numerical method based on out-of-time-order correlators (OTOCs) to identify the irreversibly decoupled environmental modes. Section~\ref{sec:dec_over} presents the computation of the decoherence functional and discusses the physical interpretation of the significance threshold. In Section~\ref{sec:unrav}, we demonstrate that our decoherent histories represent a stochastic unravelling of the exact quantum dynamics of the open system, and we present an efficient Monte Carlo simulation technique based on this insight. Finally, Sections~\ref{sec:discus} and~\ref{sec:concl} summarize our results and present conclusions.

\section{\label{sec:Dec_His} Emergent Decoherent Histories}

\subsection{Decoherence condition}

The decoherent histories approach provides a framework for understanding how classical probability rules emerge from quantum mechanics for sequences of events. In a classical description, a history is a sequence of facts occurring at discrete times $t_1 < \ldots < t_k < \ldots$. At each time $t_k$, there is a set of $m_k$ possible events, labeled by $\alpha_k = 1, \ldots, m_k$ and denoted $P_{k;\alpha_k}$. These events are mutually exclusive and exhaustive. Each event is a binary proposition: for a given history, the event that occurred is assigned the value $P_{k;\alpha_k}=1$, while all others at that time are $P_{k;\alpha'_k}=0$ for $\alpha'_k \neq \alpha_k$. The completeness of events at each time is expressed by $\sum_{\alpha_k=1}^{m_k} P_{k;\alpha_k} = 1$. A full history is therefore defined by the sequence $\boldsymbol{\alpha} = (\alpha_1, \alpha_2, \ldots)$, specifying which event happened at each time.

In the quantum formulation, each classical event $P_{k;\alpha_k}$ is represented by a projection operator $\hat{P}_{k;\alpha_k}$. The requirements of completeness and mutual exclusivity become:
\begin{align}
\sum_{\alpha_k=1}^{m_k} \hat{P}_{k;\alpha_k} &=\hat{\mathbbm{1}}\,, &
\hat{P}_{k;\alpha_k} \hat{P}_{k,\alpha'_k} &= \delta_{\alpha_k \alpha'_k} \hat{P}_{k;\alpha_k}\,.
\label{eq:mutual_ex+ex}
\end{align}
A quantum history is correspondingly represented by a time-ordered sequence of projectors interspersed with unitary evolution:
\begin{equation}
\begin{split}
\hat{h}\left(\boldsymbol{\alpha}\right)=
\ldots\hat{P}_{k;\alpha_{k}}\hat{U}\left(t_{k},t_{k-1}\right)\ldots
\hat{U}\left(t_{2},t_{1}\right)\hat{P}_{1;\alpha_{1}}\hat{U}\left(t_{1}\right) = \\
=
\ldots\hat{U}(t_k)\hat{P}_{k;\alpha_k}(t_k)\hat{P}_{{k-1};\alpha_{k-1}}(t_{k-1})\ldots\hat{P}_{1;\alpha_1}(t_1) 
\end{split}
\label{eq:quantum_history}
\end{equation}
where $\hat{U}\left(t_{k},t_{k-1}\right)$ is the evolution operator from $t_{k-1}$ to $t_k$, and $\hat{P}_{k;\alpha_k}(t_k) = \hat{U}^\dagger(t_k) \hat{P}_{k;\alpha_k} \hat{U}(t_k)$. 

For a given initial state $\ket{\Psi}$, one might define the probability of history $\boldsymbol{\alpha}$ as:
\begin{equation}
P(\boldsymbol{\alpha}) = \bra{\Psi} \hat{h}^\dagger(\boldsymbol{\alpha}) \hat{h}(\boldsymbol{\alpha}) \ket{\Psi}.
\label{eq:history_probability}
\end{equation}
However, these quantities generally fail to obey classical probability sum rules. For example, the marginal probability for an event at $t_2$ should satisfy:
\begin{equation}
P(\alpha_2) = \sum_{\alpha_1} P(\alpha_2, \alpha_1),
\notag
\end{equation}
which, using~\eqref{eq:history_probability}, translates to the quantum condition:
\begin{multline}
\bra{\Psi}\hat{U}^{\dagger}\left(t_{2}\right)\hat{P}_{2;\alpha_{2}}\hat{U}\left(t_{2}\right)\ket{\Psi}
=\\= \sum_{\alpha_{1},\alpha'_1} \bra{\Psi} \hat{P}_{1;\alpha_{1}}\left(t_{1}\right) \hat{P}_{2;\alpha_{2}}\left(t_{2}\right)\hat{P}_{1;\alpha'_1}\left(t_{1}\right)\ket{\Psi}\,.
\label{eq:history_decomposition}
\end{multline}
For this to hold, the interference between different histories must vanish. A sufficient condition for the two-time case is \cite{Griffiths:84, GMH:93, Omnes:92}:
\begin{equation*}
\Re\bra{\Psi} \hat{P}_{1;\alpha_{1}}\left(t_{1}\right) \hat{P}_{2;\alpha_{2}}\left(t_{2}\right)\hat{P}_{1;\alpha'_1}\left(t_{1}\right)\ket{\Psi} =  0\,\,
\textrm{for}\,\,\alpha_{1}\neq\alpha_{1}^{\prime}. \label{eq:two-time-consistency}
\end{equation*} 
Generalizing to arbitrary history \eqref{eq:quantum_history}, the classical probability sum rules are obeyed only if the \textit{weak decoherence} holds \cite{Griffiths:84, GMH:93, Omnes:92}:
\begin{equation}
\Re \left\langle \Psi\right|\hat{h}^{\dagger}\left(\boldsymbol{\beta}\right)\hat{h}\left(\boldsymbol{\alpha}\right)\left|\Psi\right\rangle =\delta_{\boldsymbol{\alpha}\boldsymbol{\beta}} \,P\left(\boldsymbol{\alpha}\right)\,.\label{eq:decoherence_functional}
\end{equation}

A stronger and more commonly used condition is \textit{medium decoherence} \cite{GMH:93}:
\begin{equation}
D\left(\boldsymbol{\alpha},\boldsymbol{\beta}\right)=\left\langle \Psi\right|\hat{h}^{\dagger}\left(\boldsymbol{\beta}\right)\hat{h}\left(\boldsymbol{\alpha}\right)\left|\Psi\right\rangle \approx\delta_{\boldsymbol{\alpha}\boldsymbol{\beta}} \,P\left(\boldsymbol{\alpha}\right)\,.\label{eq:decoherence_functional2}
\end{equation}
Here, $D(\boldsymbol{\alpha}, \boldsymbol{\beta})$ is the decoherence functional. Histories satisfying this condition are called consistent or decoherent.

In physical systems, exact medium decoherence is typically unattainable. Instead, one requires approximate decoherence, where off-diagonal elements are suppressed below a significance threshold: $D(\boldsymbol{\alpha}, \boldsymbol{\beta}) \approx 0$ for $\boldsymbol{\alpha} \neq \boldsymbol{\beta}$ \cite{GMH:93, Halliwell:93, Halliwell:01}.

Although medium decoherence allows one to assign probabilities to histories, it does not ensure the stability of probabilities under further refinement of the history. To address this, we adopt a stronger condition inspired by the \textit{strong decoherence} of Gell-Mann and Hartle \cite{GMH:98}. This incorporates the \textit{permanence of the past}: once recorded, information about past events remains stable and inaccessible for future interference. This physical mechanism, detailed in the next section, underlies our ``tape recorder'' model.

In practical computations, the decoherence functional is evaluated over an ensemble of histories. We define the \textit{average decoherence overlap} to quantify the overall level of consistency:
\begin{equation}
\mathcal{D} = \dfrac{1}{N^2 - N} \sum_{\boldsymbol{\alpha} \neq \boldsymbol{\beta}} \left| \bra{\Psi} \hat{h}^\dagger(\boldsymbol{\beta}) \hat{h}(\boldsymbol{\alpha}) \ket{\Psi} \right|,
\label{eq:DO}
\end{equation}
where $N$ is the number of histories. A small $\mathcal{D}$ indicates successful decoherence across the ensemble.

\subsection{Tape-recorder model as a physical mechanism \label{subsec:tape_recorder}}

We focus on the open quantum system defined by the following Hamiltonian:
\begin{equation}
\hat{H}=\hat{H}_{s}+g\hat{H}_{\rm int}+\hat{H}_{e}
\label{eq:oqs_general}
\end{equation} 
where $\hat{H}_s$ is the open quantum system, $\hat{H}_{e}$ is effectively infinite quantum environment of non-interacting quasiparticles, and $\hat{H}_{\rm int}$ is the interaction between the open system and its environment.

Our approach is guided by the tape recorder analogy illustrated in Fig.~\ref{fig:tape_recorder}. The environment acts as the tape, while the local observable functions as the read-write head. As the tape moves between reels, representing temporal evolution, the head records information onto environmental degrees of freedom. The track of recorded information emerging from under the head corresponds to an emergent decoherent history.

The head field is local, but it formally extends across the entire tape, with its influence decaying with distance. By introducing an analogous level of significance we effectively divide the environmental modes $\kappa_{p}$ (indexed by $p$) into the following regions (Fig.~\ref{fig:tape_recorder}):

\begin{figure}
\includegraphics[width=0.4\textwidth]{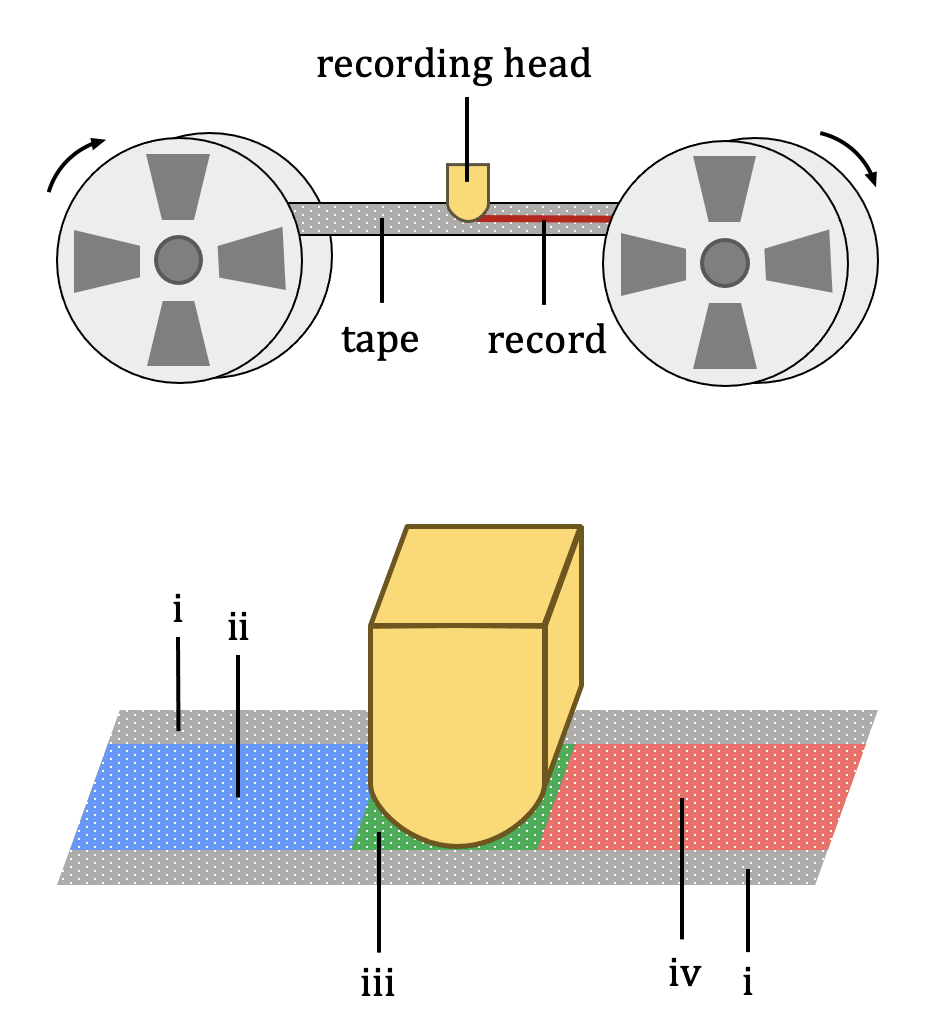}
\caption{
Tape recorder scheme. i --- non-interacting edges; ii --- future interaction regions; iii --- active recording region; iv --- irreversibly decoupled domains. In analogy with such a model, we coarse-grain the total system comprising the open system and the environment.
\label{fig:tape_recorder}}
\end{figure}

\begin{enumerate}[label=(\roman*), leftmargin=1.5em, labelwidth=1.5em, labelsep=0.2em, align=left]
\item \textit{Non-interacting edges:} Magnetic domains that never interact with the recording head, corresponding to environmental degrees of freedom $\kappa_p^{\rm edge}$
that have never interacted with the system and do not contain any information about it.

\item \textit{Future interaction regions:} Magnetic domains that will interact with the head at future arrival times $t_p^{\rm in}$, that is such degrees of freedom $\kappa_p^{\rm in}$ that after the moment $t_p^{\rm in}$ will encode system information.

\item \textit{Active recording region:} 
Magnetic domains currently interacting strongly with the head, where information is being written. These correspond to relevant environmental degrees of freedom $\kappa_p^{\rm rel}$ that currently affect system evolution.

\item \textit{Irreversibly decoupled domains:} Magnetic domains that completed recording at escape times $t_p^{\rm out}$. These correspond to irreversibly decoupled degrees of freedom $\kappa_p^{\rm out}$ that contain a stable record of the system. 

\end{enumerate}

We emphasize that this division is approximate. Just as the head formally interacts nonlocally with the entire tape, decoherent histories only approximately satisfy the consistency conditions. The significance threshold enables this practical classification, and is not a free parameter. The finite resolution of the read-write head admits a quantum analogue in the form of the experimental infidelity, which reflects unavoidable imperfections in state preparation, evolution, or measurement. Once residual coherences fall below this level, quantum superpositions become operationally indistinguishable from classical mixtures. The significance threshold therefore is determined directly by the experimentally achievable infidelity.

These conditions are formalized using the occupation number basis for the environmental modes. For each degree of freedom $\kappa_p$, we introduce the number operator $\hat{n}_{p}$ and its eigenstate projectors $\hat{P}_{p;n_p}$, where the index $n_p$ labels the occupation of mode $p$ (e.g., $n_p \in \{0, 1\}$ for a fermionic mode, or $n_p \in \mathbb{N}$ for a bosonic mode). 

The decoherence of histories is then governed by the commutators of these projectors with the interaction Hamiltonian $\hat{H}_{\rm int}(\tau)=e^{i\tau\hat{H}_{e}}\,\hat{H}_{\rm int}\,e^{-i\tau\hat{H}_{e}}$:

\vspace{\baselineskip}

For (i): $\left[\hat{H}_{\rm int}\left(\tau\right),\hat{P}_{p^{\rm edge};n_{p^{\rm edge}}}\right]\approx0,\,$ for all times; 
\vspace{\baselineskip}

For (ii): $\left[\hat{H}_{\rm int}\left(\tau\right),\hat{P}_{p^{\rm in};n_{p^{\rm in}}}\right]\approx0,\,$ for $\tau\leq t_p^{\rm in}$;

\vspace{\baselineskip}
For (iii): Projectors do not commute with the Hamiltonian for $t_p^{\rm in }<\tau<t_p^{\rm out}$; 

\vspace{\baselineskip}
For \hypertarget{ivtarget}{(iv)}: $\left[\hat{H}_{\rm int}\left(\tau\right),\hat{P}_{p^{\rm out};n_{p^{\rm out}}}\right]\approx0,\,$ for $\tau\geq t_p^{\rm out}$.

\vspace{\baselineskip}

The crucial property of the irreversibly decoupled modes (iv) is that $\hat{P}_{p^{\rm out};n_{p^{\rm out}}}$ is conserved under future evolution $[\hat{U}\left(\tau,t_p^{\rm out}\right),\hat{P}_{p^{\rm out};n_{p^{\rm out}}}]\approx0$. These modes thus serve as stable records that carry the decoherent history. As a consequence, the decoupled domains become approximate integrals of motion and the interference terms disappear:
\begin{equation}
\begin{split}
\hat{P}_{p^{\rm out};n_{p^{\rm out}}}\hat{U}^\dagger \hat{P}_{i;n_{i}}\hat{U}\hat{P}_{p^{\rm out};n_{p^{\rm out}}^{\prime}}\approx \\ \approx 
\hat{U}^\dagger\hat{P}_{i;n_{i}}\hat{P}_{p^{\rm out};n_{p^{\rm out}}}\hat{P}_{p^{\rm out};n_{p^{\rm out}}^{\prime}}\hat{U}\approx\\ \approx\delta_{n_{p^{\rm out}}n_{p^{\rm out}}^{\prime}}\hat{U}^\dagger\hat{P}_{i;n_{i}}\hat{P}_{p^{\rm out};n_{p^{\rm out}}}\hat{U}   
\end{split}
\end{equation}
where $\hat{U} \equiv\hat{U}\left(\tau,t_p^{\rm out}\right)$, $t_i>t_p^{\rm out}$ and we have employed the mutual-exclusiveness of projections (\ref{eq:mutual_ex+ex}). In this way, we perform a coarse-graining of the environment. The existence of irreversibly decoupled modes constitutes our strong decoherence condition, which determines the decoherent histories. 

Our mechanism of decoherence arises from the local structure of the interaction and the presence of irreversibly decoupled modes, which encode the history in a physically meaningful and controllable way.

In the following sections we present an efficient procedure how to construct the stream of projectors $\hat{P}_{1^{\rm out};n_{1^{\rm out}}},\hat{P}_{2^{\rm out};n_{2^{\rm out}}},\ldots\hat{P}_{p^{\rm out};n_{p^{\rm out}}},\ldots$ and the corresponding times $t_p^{\rm out}$ for a specific type of model (\ref{eq:oqs_general}) based on the principal component analysis of the interaction OTOC \cite{PolAref:24}.

\section{\label{sec:model}Considered physical system} 

Although our considerations are general, we now apply the tape-recorder model of decoherent histories to a specific physical system. We treat the open system together with its environment as a large closed system. Within this framework, we introduce a model of a local open quantum system and define a local interaction quench (hereinafter, the natural system of units is used everywhere: $\hbar=1$).

\subsection{Local interaction quench}
The open quantum system model we consider is a specific type of the model (\ref{eq:oqs_general}):
\begin{equation}
\hat{H}=\hat{H}_{s}+g\hat{V}_{s}^{\dagger}\hat{a}_{0}+g\hat{V}_{s}\hat{a}_{0}^{\dagger}+\hat{H}_{e}\label{eq:star_oqs_model}
\end{equation}
where $\hat{H}_{\rm int}=\hat{V}_{s}^{\dagger}\hat{a}_{0}+\hat{V}_{s}\hat{a}_{0}^{\dagger}$ is the bilinear coupling between the open system via $\hat{V}_{s}$ and the environment via $\hat{a}_{0}$. The environment is defined through its normal mode decomposition:
\begin{equation}
\hat{H}_{e}=\int\limits_0^\infty d\omega\, \omega \,\hat{a}^{\dagger}\left(\omega \right)\,\hat{a}\left(\omega \right)  \label{eq:star_environment_representation}
\end{equation}
where $\left[\hat{a}\left(\omega\right),\hat{a}^{\dagger}\left(\omega^{\prime}\right)\right]_{\pm}=\delta\left(\omega-\omega^{\prime}\right)$, with $\left[\cdot,\cdot\right]_{\mp}$ being a commutator/anticommutator in the case of bosonic/fermionic environment. The coupling site $\hat{a}_{0}$ is expanded in terms of the normal modes as:
\begin{equation}
\hat{a}_{0} = \int\limits_0^\infty d\omega\,c\left(\omega\right)\,\hat{a}\left(\omega\right)\label{eq:star_coupling_site}
\end{equation}
here the spectral density of states of the environment is $J\left(\omega\right) = \left|c\left(\omega\right)\right|^{2}$ and $\int\limits_0^\infty d\omega \,J\left(\omega\right)=1$. It is well known that the influence of the bath on the system is entirely characterized by its spectral density in case of linear environment \cite{Chin:10}.

In the interaction picture with respect to the free environment, the Hamiltonian becomes: 
\begin{equation}
\hat{H}\left(t\right)=\hat{H}_{s}+g\hat{V}_{s}^{\dagger}\hat{a}_{0}\left(t\right)+g\hat{V}_{s}\hat{a}_{0}^{\dagger}\left(t\right)\label{eq:oqs_interaction_pic-1}
\end{equation}
with $\hat{a}_{0}\left(t\right)=\int\limits_0^\infty d\omega\, c\left(\omega \right)\hat{a}\left(\omega\right)e^{-i\omega t}$.

For our purposes, it is convenient to transform the representation bosonic bath from normal mode (\ref{eq:star_environment_representation}) to the equivalent chain representation. This is achieved via a unitary transformation $\hat{U}_n\left(\omega\right)$ \cite{Chin:10}:
\begin{equation}
\hat{H}_{e}=\sum_{n=0}^{\infty}\left\{ \varepsilon_{n}\hat{a}_{n}^{\dagger}\hat{a}_{n}+h_{n}\hat{a}_{n}^{\dagger}\hat{a}_{n+1}+h_{n}\hat{a}_{n+1}^{\dagger}\hat{a}_{n}\right\} \label{eq:semiinf_chain}
\end{equation}
where $\hat{a}_n = \int\limits_0^\infty d\omega\,\hat{U}_n\left(\omega\right)\,\hat{a}\left(\omega\right)$, $\left[\hat{a}_{k},\hat{a}_{l}^{\dagger}\right]_{\pm}=\delta_{kl}$  and the coupling site operator $\hat{a}_{0}$
appears as the first site of the chain.

It is important to emphasize that the integrability of the total system depends on the choice of the open-system Hamiltonian $\hat{H}_{s}$. For example, if one selects a single site of the environment as the open system, the total system becomes integrable, and decoherent histories can still be constructed in a similar manner (as described below). Thus, our mechanism does not rely on whether the open system itself is integrable or non-integrable.

\begin{figure}
\includegraphics[width=0.48\textwidth]{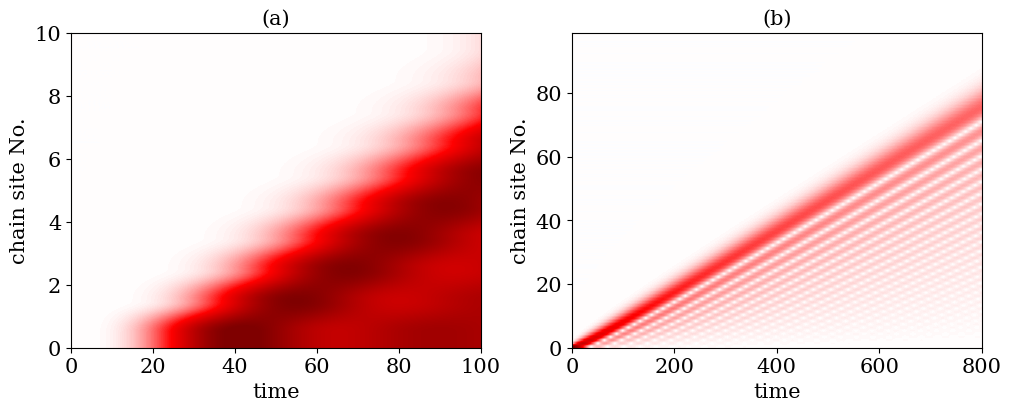}
\caption{\label{fig:local_quench}
When local open system is coupled to the environment at $t=0$, the disturbance propagates in a light-cone-like way. In (a) $\hat{n}_{p}\left(t\right)=\bra{\Psi\left(t\right)}\hat{a}_{p}^{\dagger}\hat{a}_{p}\ket{\Psi\left(t\right)}$ is plotted depending on chain site over time. In (b) the light-cone-like spread of the coupling site $\hat{a}_{0}^{\dagger}\left(t\right)$, we plot $\left|\alpha_{k}\left(t\right)\right|$, eq. (\ref{eq:spread_equation}), with respect to the chain site index $k$ and time $t$.}
\end{figure} 

Suppose that initially the joint state of open system and environment is a pure product state: 
\begin{equation}
\ket{\Psi\left(0\right)} = \ket{\phi} _{s}\otimes\ket{0}_{e}
\label{eq:initial_cond}
\end{equation}
where $\ket{\phi}_{s}$ is some state inside the Hilbert space of open system, $\ket{0}_{e}$ is vacuum state inside Fock space of environment. After $t=0$ it evolves according to Schrodinger equation:
\begin{equation}
i\partial_{t}\ket{\Psi\left(t\right)} =\hat{H}\left(t\right)\ket{\Psi\left(t\right)}
\label{eq:interaction_pic_sch_eq}
\end{equation}
with $\hat{H}\left(t\right)$ given by eq.~\eqref{eq:oqs_interaction_pic-1}. This is called the local interaction quench. It leads to a light-cone-like spread of quasiparticles inside the environment. According to the Lieb-Robinson bounds \cite{Lieb-Rob1972}, the front of the light cone propagates along the chain~\eqref{eq:semiinf_chain} with some finite velocity $v_{L}$. 

The local interaction quench is accompanied with a light-cone-like spread of operator $\hat{a}_{0}^{\dagger}\left(t\right)$:
\begin{equation}
\hat{a}_{0}^{\dagger}(t) = \sum_{k=0}^{\infty} \alpha_{k}(t) \hat{a}_{k}^{\dagger}
\label{eq:spread_equation}
\end{equation}
where $\alpha_{k}\left(t\right)$ obeys to a single-particle first-quantized Schrodinger equation of the free environment:
\begin{equation}
i\partial_{t}\alpha_{k}\left(t\right)=\varepsilon_{k}\alpha_{k}\left(t\right)+h_{k}\alpha_{k+1}\left(t\right)+h_{k-1}\alpha_{k-1}\left(t\right) \label{eq:a0_spread_eq}
\end{equation}
with $h_{-1}\equiv0$ and the initial condition is: 
\begin{equation}
\alpha_{k}\left(0\right)=\delta_{k0} \label{eq:a0_spread_eq_ini_cond}
\end{equation}

The coefficients $\alpha_k(t)$ can be interpreted as a one particle quantum state $\ket{\alpha(t)}$:
\begin{equation}
\ket{\alpha(t)} = \hat{a}^\dagger_0(t) \ket{0}_e = \sum_{k=0}^{\infty} \alpha_{k}(t) \ket{k}
\end{equation}
with the site amplitude $\langle\kappa|\alpha(t)\rangle = \alpha_k(t)$, here $\ket{k}$ is quantum localized in $k$ chain site.

\begin{figure}[h!]
\centering
\includegraphics[width=0.4\textwidth]{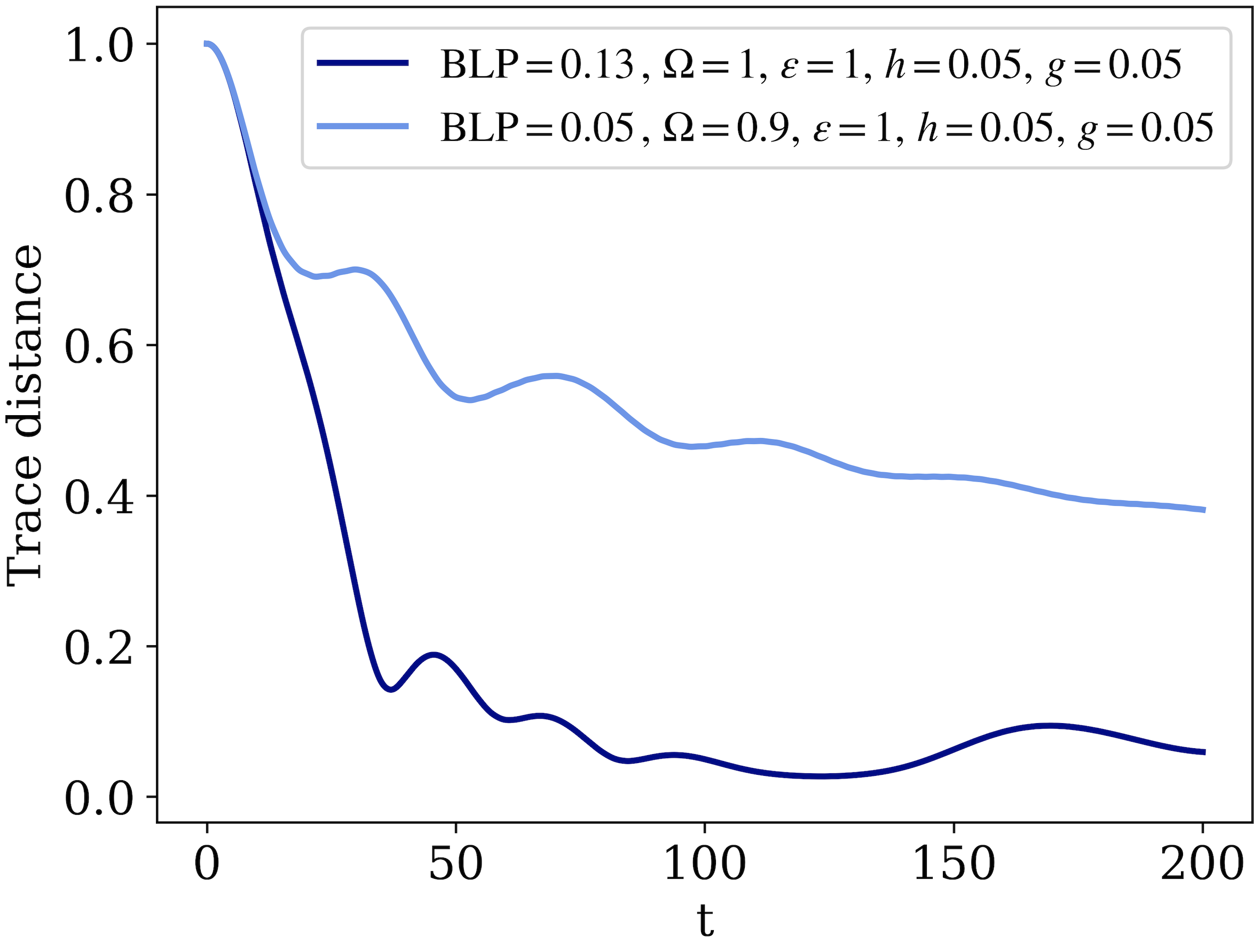}
\hfill
\includegraphics[width=0.4\textwidth]{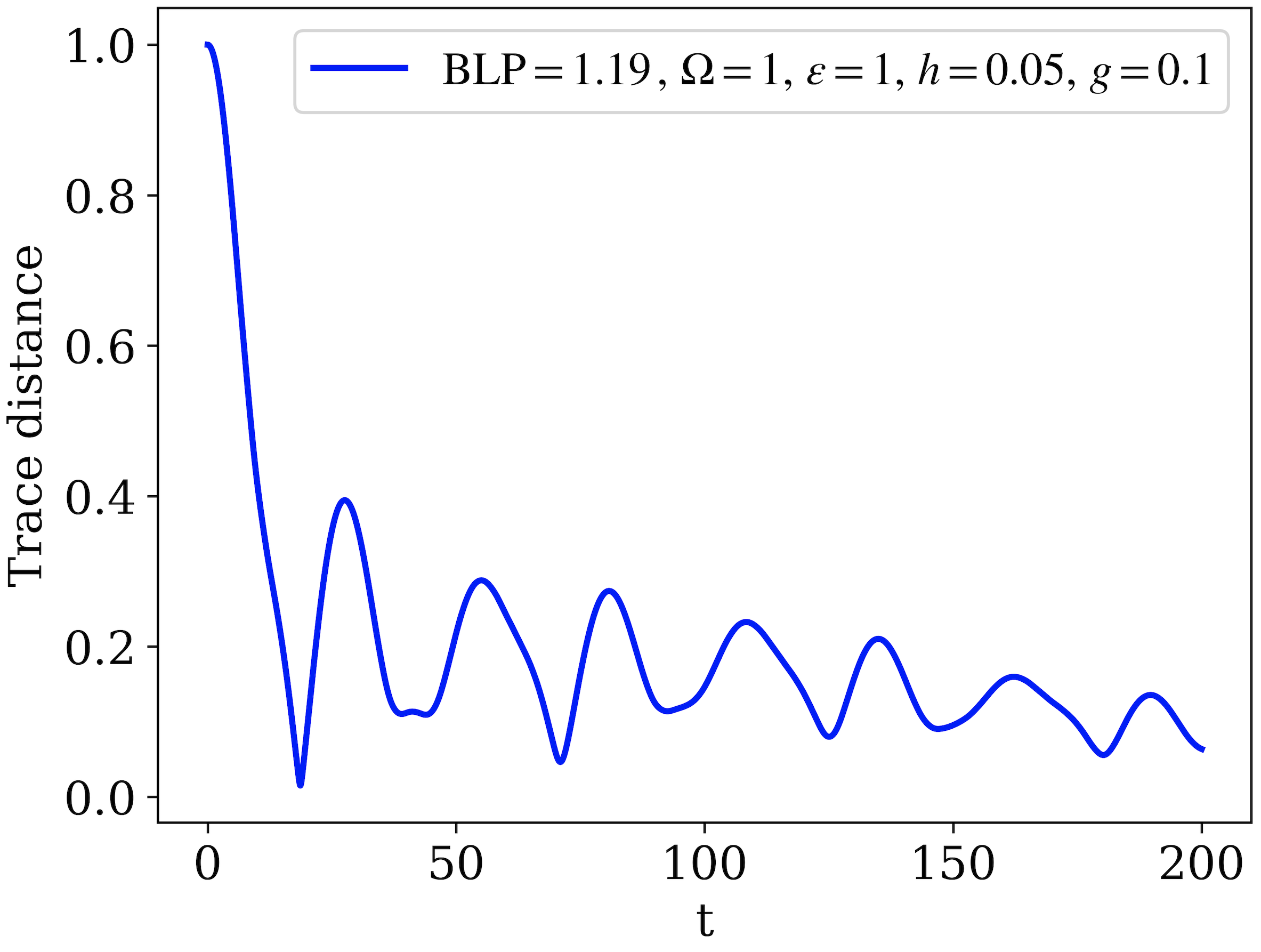}
\caption{Time dependence for the trace distance $T(\rho_1,\rho_2; t) = \dfrac{1}{2}\sqrt{(\rho_1(t) - \rho_2(t))^2}$ for different parameters of model (\ref{eq:oqs_interaction_pic-1}, \ref{eq:semiinf_chain}).
For coupling $g=0.05$ and a qubit frequency at the center of the environmental band ($\Omega=1$), weak oscillations are observed, with a BLP measure of $0.13$. 
Shifting the qubit frequency toward the band edge ($\Omega=0.9$) makes the oscillations more visible, while the BLP value is $0.05$. 
For stronger coupling ($g=0.1$), large-amplitude oscillations appear and the BLP measure increases to $1.19$, indicating a strongly non-Markovian regime. 
As the BLP is a scalar quantity, the full time dependence of the trace distance provides a more complete characterization of the dynamics.
\label{fig:BLP}}
\end{figure}

As an example, in Fig.~\ref{fig:local_quench} (a) we provide a numerically exact calculation of chain site occupations for the case when the open system is a driven qubit: $\hat{H}_{s}\left(t\right)=\Omega \,\hat{\sigma}_{+}\hat{\sigma}_{-}+\hat{\sigma}_{x}\,f\cos t$ and the environment is bosonic. Here $f = 0.1$ and other parameters of model eq. (\ref{eq:oqs_interaction_pic-1}, \ref{eq:semiinf_chain}) are: $\Omega = 1$, $g=0.05$, $\hat{V}_{s}=\hat{\sigma}_{-}$, $\varepsilon_{j}\equiv1$, $h_{j}\equiv0.05$. In Fig.~\ref{fig:local_quench} (b) we illustrate the light-cone-like spread of $\hat{a}_{0}^{\dagger}\left(t\right)$ by plotting $\left|\alpha_{k}\left(t\right)\right|$ as a function of chain site number and time with maximum Lieb-Robinson velocity $v_L = 2\,h_j$.

The dynamics in this model is generally non-Markovian. To quantify non-Markovianity we evaluate the time-dependent trace distance $T(\rho_1,\rho_2; t) = \dfrac{1}{2}\sqrt{(\rho_1(t) - \rho_2(t))^2}$ for two orthogonal initial qubit states $\rho_1(0) = \ket{\uparrow}\bra{\uparrow}$ and $\rho_2(0) = \ket{\downarrow}\bra{\downarrow}$. The trace distance exhibits clear non-monotonic behavior, indicating non-Markovian dynamics and information backflow from the environment to the system \cite{Breuer2009} (see Fig.~\ref{fig:BLP}).

\subsection{``Tape recorder'' coarse-graining}
Different magnetic domains can be understood as environmental degrees of freedom that lie inside (or outside) the light cone generated by the local interaction quench (\ref{eq:spread_equation}). Using the tape recorder analogy (see Sec.~\ref{subsec:tape_recorder}), we provide a coarse-grained description of the system by identifying (i) never interacting, (ii) about to interact, (iii) currently interacting (inside the quench light cone), and (iv) already interacted and irreversibly decoupled degrees of freedom, which support decoherent histories. 

\begin{figure}
    \centering
    \includegraphics[width=0.5\textwidth]{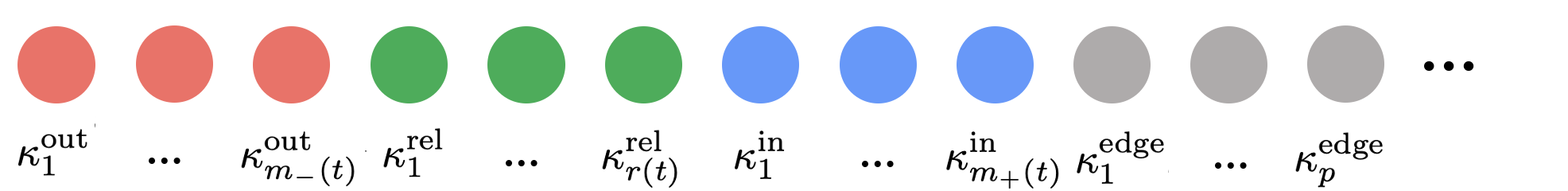}
    \caption{Mapping of the environmental modes onto a ``moving tape''. At time $t$, the tape is divided into four regions: edge, incoming ($m_+(t)$ modes), relevant ($r(t)$ modes), and outgoing domains ($m_-(t)$ modes), where $r(t) = m_+(t) - m_-(t)$.}
    \label{fig:bogol_trs}
\end{figure}

To realize this four-region coarse-graining, we transform the original chain modes $\hat{a}_{0}^{\dagger}, \hat{a}_{1}^{\dagger}, \ldots$ into new modes corresponding to the tape recorder domains. In this way, the environment is mapped onto a ``moving tape''. This is achieved by a time-dependent Bogoliubov transformation, which naturally generates wave-packet–like localized modes:
\begin{equation}
\hat{\kappa}_{p}^{\dagger}(t)=\sum_{q=0}^{\infty}U_{pq}(t)\hat{a}_{q}^{\dagger}
\label{eq:trial_DoF}
\end{equation}
with $\left[UU^{\dagger}\right]_{pq}=\delta_{pq}$. We introduce the corresponding one particle quantum state:
\begin{equation}
\ket{\kappa_p} = \hat{\kappa}_p^\dagger\,\ket{0}_e = \sum_{q=0}^{\infty}U_{pq}(t)\ket{q} 
\end{equation}
where $\ket{q}$ is quantum localized in $q$ chain site.
We refer these modes as follows (i) $\kappa_p^{\rm edge}$, (ii) $\kappa_p^{\rm in}$, (iii) $\kappa_p^{\rm rel}$, (iv) $\kappa_p^{\rm out}$ (Fig.~\ref{fig:bogol_trs}).

At a given time, all modes are linearly independent and an effective Hamiltonian only with $\kappa_p^{\rm rel}$ appears in this moving basis (is obtained in the following section). 

Due to such transformation the irreversibly decoupled modes are defined as those whose projections commute with the Hamiltonian (\hyperlink{ivtarget}{iv}) after some escape time. These modes then act as stable records, irreversibly storing information about the system and thereby defining decoherent histories.

\section{\label{sec:comp_fram} Computational Framework}

In this section we will explicitly derive environmental degrees of freedom, which carry significant information about open system, together with the corresponding time moments. We show that the sequence of projections onto its occupation number eigensubspace after escape times satisfy the approximate decoherence condition with increasing accuracy for decreasing significance threshold. 

The problem of finding decoherent projections is reduced to the problem of finding the appropriate Bogolubov transformation. We need to consider the quench light cone (\ref{eq:spread_equation}), not only in the usual spatial frame, but in an arbitrary rotated one (\ref{eq:trial_DoF}) and also need to estimate the time moments when the mode $\kappa_{p}$ arrives inside the light cone, and the time it escapes the light cone. This is done by principal component analysis of the interaction out-of-time-ordered correlator (OTOC).

\subsection{Irrelevant subspace for local interaction quench: edges of tape \label{subsec:Lieb-Robinson-metric}}
Let's start the construction of the Bogoliubov transformation by identifying the degrees of freedom corresponding to the edges of the tape (i), i.e., modes with which the open system never significantly interacts. If the recording lasts up to a time $T$, then the edges of the tape are precisely those modes into which no information is ever written during $[0,T]$.

Suppose we are given some trial degree of freedom $\hat{\kappa}^{\dagger}$  (\ref{eq:trial_DoF}). For $\hat{\kappa}^{\dagger}$ to lie outside the quench light cone by the time $T$, it must not interact significantly with the open system over the time interval $[0, T]$. This condition can be formalized in terms of the following OTOC \cite{PolAref:24}:
\begin{equation}
\mathcal{C}\left(\kappa,t\right)={}_e \bra{0}\left[\hat{a}_{0}\left(t\right),\hat{\kappa}^{\dagger}\right] \left[\hat{a}_{0}\left(t\right),\hat{\kappa}^{\dagger}\right]^\dagger\ket{0}_e
\end{equation}
which yields an estimate of the instant interaction intensity with the open system at time $t$. The light cone is governed by the time-averaged, rather than the instantaneous, interaction strength of the mode. To identify $\hat{\kappa}^{\dagger}{}^{\rm edge}$ as irrelevant for the system dynamics, the average intensity $\,\mathcal{I}^{+}\left(\kappa,T\right)=\int\limits_{0}^{T}d\tau \,\,\mathcal{C} \left(\kappa,\tau\right)\,$ should remain below a certain significance threshold $a_{\rm cut}$:

\begin{equation}
g_{+}\left(\kappa,T\right) = \mathcal{I}^{+}\left(\kappa,T\right) -a_{\rm cut}<0\,.
\label{eq:LiebRobinsonMetriс}
\end{equation}

This condition can be conveniently rewritten using the state $|\kappa\rangle$. Indeed, by introducing the retarded light-cone density matrix:

\begin{equation}
\hat{\rho}_{+}\left(T\right)=\intop_{0}^{T}d\tau\left|\alpha\left(\tau\right)\right\rangle \left\langle \alpha\left(\tau\right)\right| \label{eq:rho_plus}
\end{equation}
with $\alpha\left(\tau\right)$ being the solution of the operator spread equations (\ref{eq:a0_spread_eq}–\ref{eq:a0_spread_eq_ini_cond}) and using the relation:

\begin{equation}
\begin{gathered}
\left[\hat{a}_{0}\left(\tau\right),\hat{\kappa}^{\dagger}\right] = \left[\sum_{q=0}^\infty\alpha_q^*(\tau)\hat{a}_q,\sum_{p=0}^\infty\kappa_p\hat{a}_p^{\dagger}\right]=\\
=\sum_{p=0}^\infty \alpha_p^*(\tau) \kappa_p = \langle\alpha(\tau)|\kappa\rangle 
\end{gathered}
\end{equation}
which follows from (\ref{eq:spread_equation}) and (\ref{eq:trial_DoF}), the average intensity takes the form:
\[
\begin{gathered}
\mathcal{I}^{+}\left(\kappa,T\right) = {}_e \bra{0}\int\limits_{0}^{T}d\tau \left[\hat{a}_{0}\left(\tau\right),\hat{\kappa}^{\dagger}\right] \left[\hat{a}_{0}\left(\tau\right),\hat{\kappa}^{\dagger}\right]^\dagger\ket{0}_e =
\\
= \left\langle \kappa\right|\hat{\rho}_{+}\left(T\right)\left|\kappa\right\rangle \,.
\end{gathered}
\]
Consequently, the condition (\ref{eq:LiebRobinsonMetriс}) becomes:
\begin{equation}
g_{+}\left(\kappa,T\right)= \left\langle \kappa\right|\hat{\rho}_{+}\left(T\right)\left|\kappa\right\rangle - a_{\rm cut} < 0\,.
\label{eq:Lieb_robinson_metric}
\end{equation}

Observe that the above definition of light cone is consistent: 
\begin{equation}
\textrm{if}\,\,g_{+}\left(\kappa,t\right)<0\,\,\textrm{then}\,\,g_{+}\left(\kappa,t^{\prime}\right)<0\,\,\textrm{for all}\,\,t^{\prime}<t\label{eq:light_cone_consistency}
\end{equation}
which follows from (\ref{eq:rho_plus}). 

The state with maximal average intensity $\mathcal{I}^{+}\left(\kappa,t\right)=\left\langle\kappa\right|\hat{\rho}_{+}\left(t\right)\left|\kappa\right\rangle$, that is the most significant light-cone mode over $\left[0,t\right]$ is:
\begin{equation}
\left|\widetilde{\kappa}_{1}\right\rangle =\arg\max\frac{\left\langle \kappa\right|\hat{\rho}_{+}\left(t\right)\left|\kappa\right\rangle }{\left\Vert \kappa\right\Vert ^{2}}\,.
\end{equation}
According to the Ritz variational principle, the solution is given by eigenstate for the largest eigenvalue of $\hat{\rho}_{+}\left(t\right)$:
\begin{equation}
\hat{\rho}_{+}\left(t\right)\left|\widetilde{\kappa}_{1}\right\rangle =\mathcal{I}_{1}^{+}\left(t\right)\left|\widetilde{\kappa}_{1}\right\rangle
\label{eq:rho_plus_eigvals}
\end{equation}
where $\mathcal{I}_{1}^{+}\left(t\right)\equiv\mathcal{I}^{+}\left(\widetilde{\kappa}_{1},t\right)$. The threshold $a_{\rm cut}$ can be specified by choosing a relative threshold $r_{\rm cut}$ (e.g. $r_{\rm cut} = 10^{-4}$):
\begin{equation}
a_{\rm cut} = r_{\rm cut} \, \mathcal{I}_{1}^{+}\left(t\right)\,.
\label{eq:a_cut}
\end{equation}
The subspace of irrelevant modes $\ket{\kappa^{\rm edge}}$ correspond to the edges of tape $\ket{\kappa^{\rm edge}}$ is defined through:
\begin{equation}
g_{+}\left(\kappa,T\right)=\mathcal{I}^{+}\left(T\right)-r_{\rm cut}\,\mathcal{I}_{1}^{+}\left(T\right) <0\,.
\label{eq:cond_edge}
\end{equation}

They contribute negligibly to the full many-body evolution. This is quantified via the square-root fidelity:
\begin{equation}
\sqrt{F(t)}=\left|
\langle\Psi_{\perp}(t)|\Psi(t)\rangle\right|
\label{eq:sqrt_fidelity}
\end{equation}
where $\left|\Psi_{\perp}\left(t\right)\right\rangle $ evolves under $\hat{H}_{\perp}\left(t\right)$ obtained from $\hat{H}\left(t\right)$ by discarding $\hat{\kappa}^{\rm edge},\,\hat{\kappa}^{\dagger}{}^{\rm edge}$ (see Appendix \ref{app:just_Lieb_Rob}).
The corresponding infidelity: 
\begin{equation}
I\left(t\right)=1-\sqrt{F\left(t\right)}\label{eq:infidelity}
\end{equation}
yields a measure of the significance of $\kappa$ during $\left[0,t\right]$ and converges to zero for modes outside light-cone interior, corresponding the edges of tape.

The interior of the light cone consists only of modes whose statistical contribution is not negligible relative to the maximum. The eigenstates $\ket{\widetilde{\kappa}_p}$, sorted in the decreasing order of eigenvalues $\mathcal{I}_{p}^{+}\left(t\right)$, are the basis of the most significant degrees of freedom for the interaction with the open system during time interval $\left[0,t\right]$. For the state $\left|\widetilde{\kappa}_p\right\rangle$ the light cone interior condition is:
\begin{equation}
g_{+}\left(\kappa,t\right)=\mathcal{I}^{+}\left(t\right)-r_{\rm cut}\,\mathcal{I}_{1}^{+}\left(t\right) >0\,.\label{eq:normal_mode_condition}
\end{equation}
There is a finite number $m\left(T\right)$ of states $\ket{\widetilde{\kappa}_1} \ldots|\widetilde{\kappa}_{m\left(T\right)}\rangle$ which satisfy it on $\left[0,T\right]$. They correspond to the region of tape which record information. We call them the \textit{light cone interior normal modes}.

If we consider the analogy with the Lieb–Robinson light cone in mode space and include $m\left(t\right)=Bt+\delta$ light cone interior normal modes, the resulting infidelity is: 
\begin{equation}
I\left(t\right)\lesssim\left[g\left\Vert \hat{V}_{s}\right\Vert \right]^{2}\mathcal{I}_{1}^{+}\left(\infty\right)e^{-\gamma\delta}
\label{eq:rel_guard_modes_convergence}
\end{equation}
and does not depend on time, providing uniform exponential convergence with respect to the number $\delta$ of additional ``guard'' modes. Here $B$, $\gamma$ is some positive constant (see Appendix \ref{app:just_Lieb_Rob}). Equivalently, the choice $\delta$ corresponds to a relative cutoff $r_{\rm cut}=e^{-\gamma\delta}$. 

\subsection{Basis of a minimal light cone normal modes \label{subsec:Frame-of-a-minimal-lc}}

Above we defined a subspace of significant modes (\ref{eq:normal_mode_condition}) where information about open system is recorded (Fig. \ref{fig:tape_recorder}, blue and green regions (ii, iii)).

For each such degree of freedom there is an arrival time $t_{\rm in}\left(\kappa\right)$ when it enters the light cone and become significant. Formally it can be described as follows:
\begin{equation}
t_{\rm in}\left(\kappa\right)=\int\limits_{0}^{T}d\tau\,\theta\left[\,-\,g_{+}\left(\kappa,\tau\right)\right]
\label{eq:arrival_time}
\end{equation}
where $\theta\left[x\right]$ is a Heaviside step function.

\begin{figure}[h!]
\centering
\includegraphics[width=0.23\textwidth]{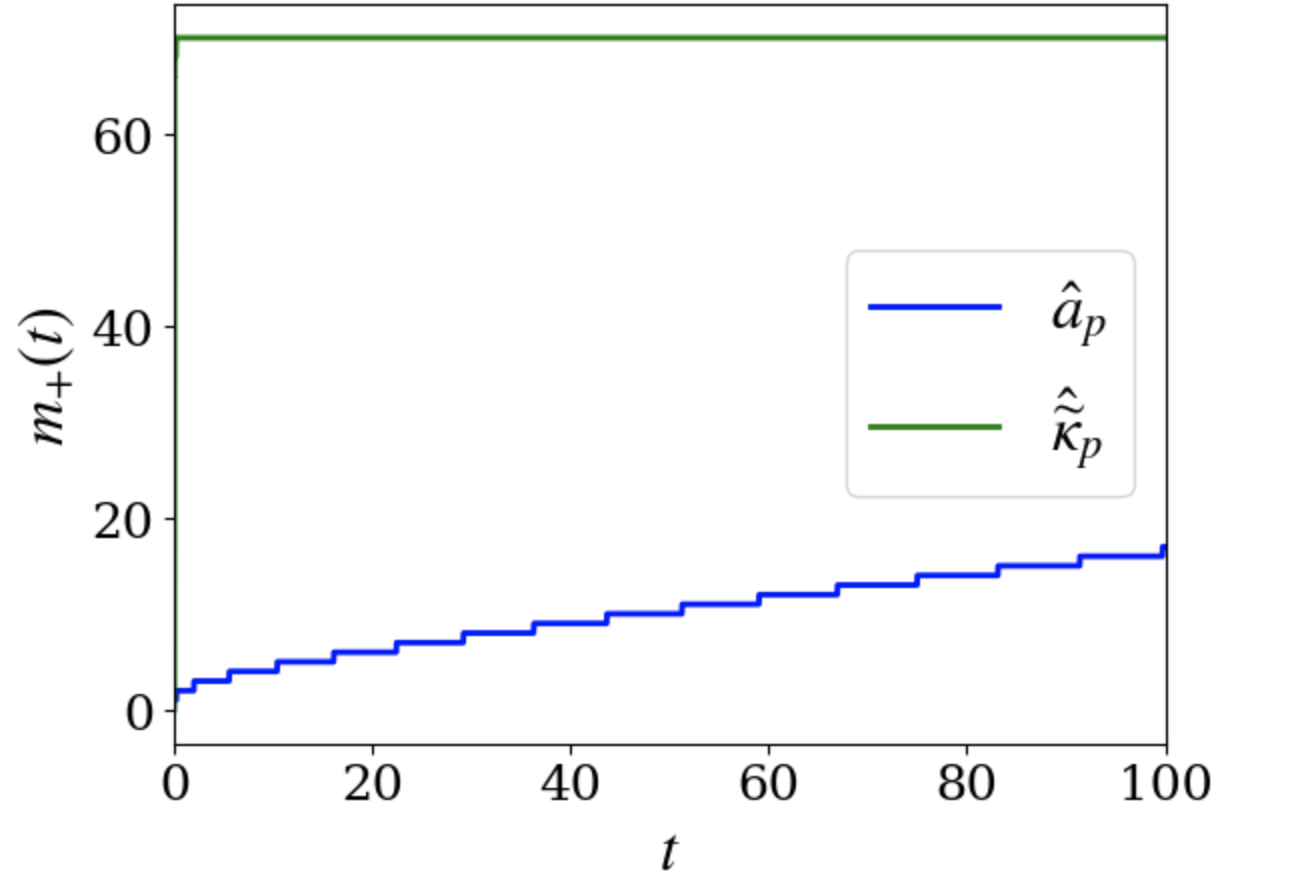}
\hfill
\includegraphics[width=0.23\textwidth]{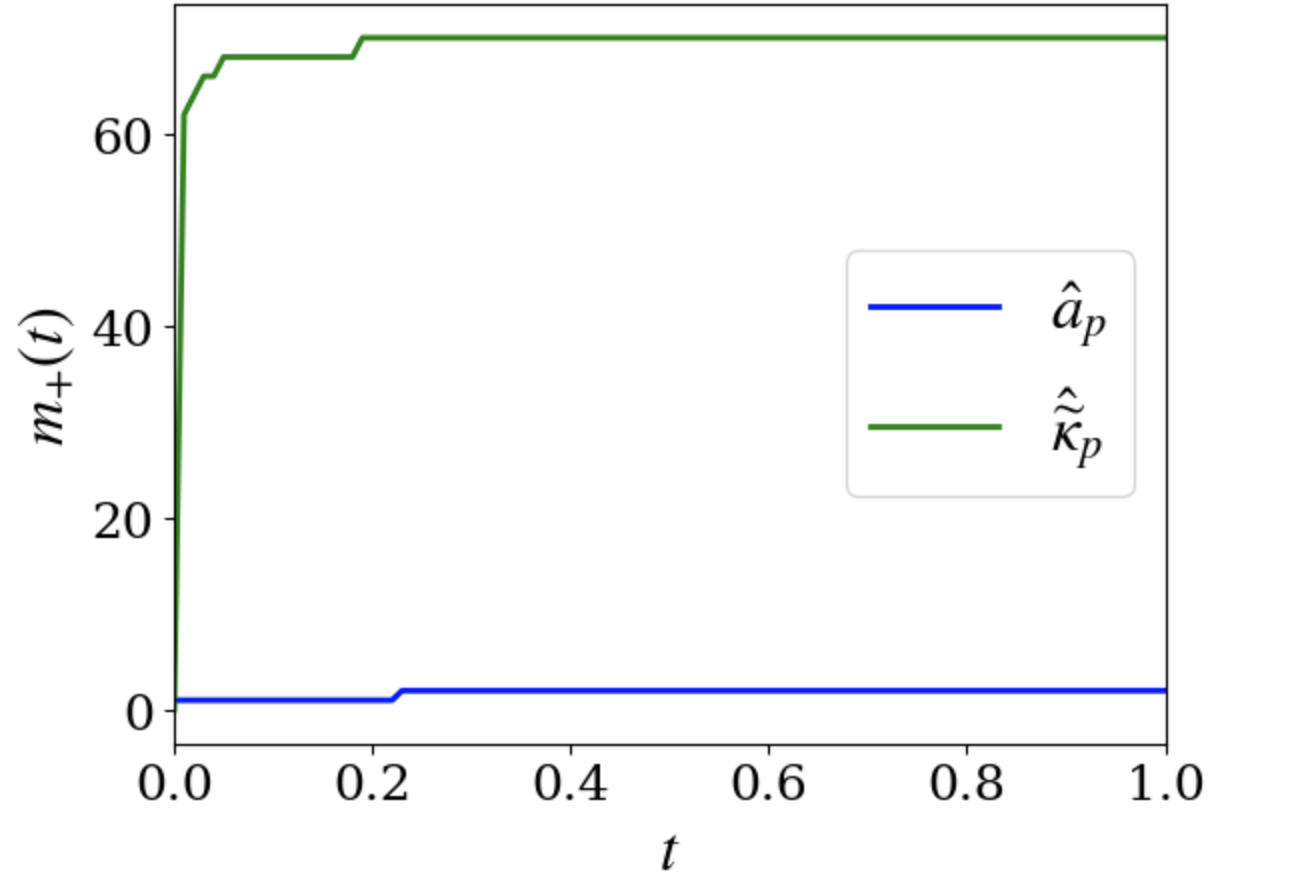}
\caption{The number of arrived modes $m_{+}(t)$ as a function of time $t$ is shown for different choices of basis.
Here $m_{+}(t)$ is defined as the number of modes $\kappa$ such that $t_{\rm in}(\kappa) \le t$.
The parameters used are: $a_{\rm cut}=10^{-5}$, $\varepsilon_j \equiv 1$, $h_j \equiv 0.05$.
The behavior of $m_{+}(t)$ is strongly basis dependent.
In the chain basis, the modes couple one-by-one at an asymptotically constant rate (blue curve). In contrast, in the normal-mode basis corresponding to the light-cone interior normal modes, the modes couple to the system almost instantaneously at $t\approx0$ (green curve).}
\label{fig:arrival_time}
\end{figure}

In different bases, the rates at which the modes enter the light cone are completely different. In the chain basis, the light cone is well-defined, but the modes themselves are not statistically independent. At the same time, in the basis of normal modes, the operator (\ref{eq:spread_equation}) spreads almost instantly over the entire system, making the light cone essentially absent.
In Fig. \ref{fig:arrival_time} we present the arrival times for chain sites $\hat{a}_{p}$
and for light cone interior normal modes $\hat{\widetilde{\kappa}}_p$.

Following the tape recorder analogy (Sec. \ref{subsec:tape_recorder}), we retain only the modes that carry statistically significant records (i.e., the light-cone interior normal modes) and that also couple sequentially, as in the chain sites. To optimize the localization of independent degrees of freedom, we rotate the normal mode basis:
\begin{equation}
\hat{\bar\kappa}_{p}^{\dagger} = \sum_{q=1}^{m(T)} U_{pq} \hat{\widetilde{\kappa}}_q^{\dagger}, \quad p,q = 1, \ldots, m(T),
\end{equation}
with $U U^{\dagger} = \mathbbm{1}$. Each operator $\hat{\bar\kappa}_{p}^{\dagger}$ is associated with an arrival time $t_{\rm in}(\bar\kappa_p)$ (Eq. \ref{eq:arrival_time}), and the operators are sorted in order of increasing arrival times. There exists a unique \textit{minimal-light-cone} basis, in which the degrees of freedom remain as localized as possible outside the light cone.

The unitary rotation which gives the minimal light cone basis is \cite{Pol2022}:
\begin{equation}
\hat{\kappa}_{p}^{\rm in \dagger}=\sum_{q=1}^{m\left(T\right)}\left[U^{+ \rm min}\right]_{pq}\,\hat{\widetilde{\kappa}}_q^{\dagger}
\end{equation}
and $\hat{\kappa}_{p}^{\rm in \dagger}$ are such modes, whose arrival times $t_{\rm in}\left(\kappa_{p}^{\rm in}\right)$ are delayed as much as possible. Details of the construction are given in the Appendix~\ref{app:find_MFLCB} and \cite{Pol2022}. These modes contain significant information about the open system, the decoherent history is recorded in them (ii).  

Thus, on a finite time interval $\left[0,T\right]$ the full system dynamics (\ref{eq:oqs_interaction_pic-1}) can be approximated by effective Hamiltonian $H_{+ \rm min}(t)$ in the frame of the minimal light cone only with $m_+(t)$ light cone modes (Appendix \ref{app:relevant_sub}):
\begin{equation}
\begin{split}
\hat{H}\left(t\right)\approx\hat{H}_{+ \rm min}\left(t\right)=\hat{H}_{s}+g\,\hat{V}_{s}^{\dagger}\sum_{k=1}^{m_{+}\left(t\right)}\chi_{k}^{\rm in }{}^{*}\left(t\right)\hat{\kappa}_{k}^{\rm in}\,+\\+\,g\,\hat{V}_{s}\sum_{k=1}^{m_{+}\left(t\right)}\chi_{k}^{\rm in}\left(t\right)\hat{\kappa}_{k}^{\rm in \dagger}
\end{split}
\label{eq:H_min}
\end{equation}
with an accuracy of error controlled by the infidelity bound (\ref{eq:rel_guard_modes_convergence}). 

\noindent Here $\chi_{k}^{\rm in}\left(t\right)=\sum\limits_{p=1}^{m\left(T\right)}\left[U^{+ \rm min}\right]_{pk}^{*}\left\langle \widetilde{\kappa}_{p}\left|\alpha\left(t\right)\right.\right\rangle$ and $m_{+}\left(t\right)$ is the number of modes which have arrived before $t$:
\begin{equation}
m_{+}\left(t\right)=\intop_{0}^{t}d\tau\sum_{p=1}^{m\left(T\right)}\delta\left(\tau-t_{\rm in}\left(\kappa_{p}^{\rm in}\right)\right)
\label{eq:m_plus}
\end{equation}
\begin{figure}
\includegraphics[width=0.4\textwidth]{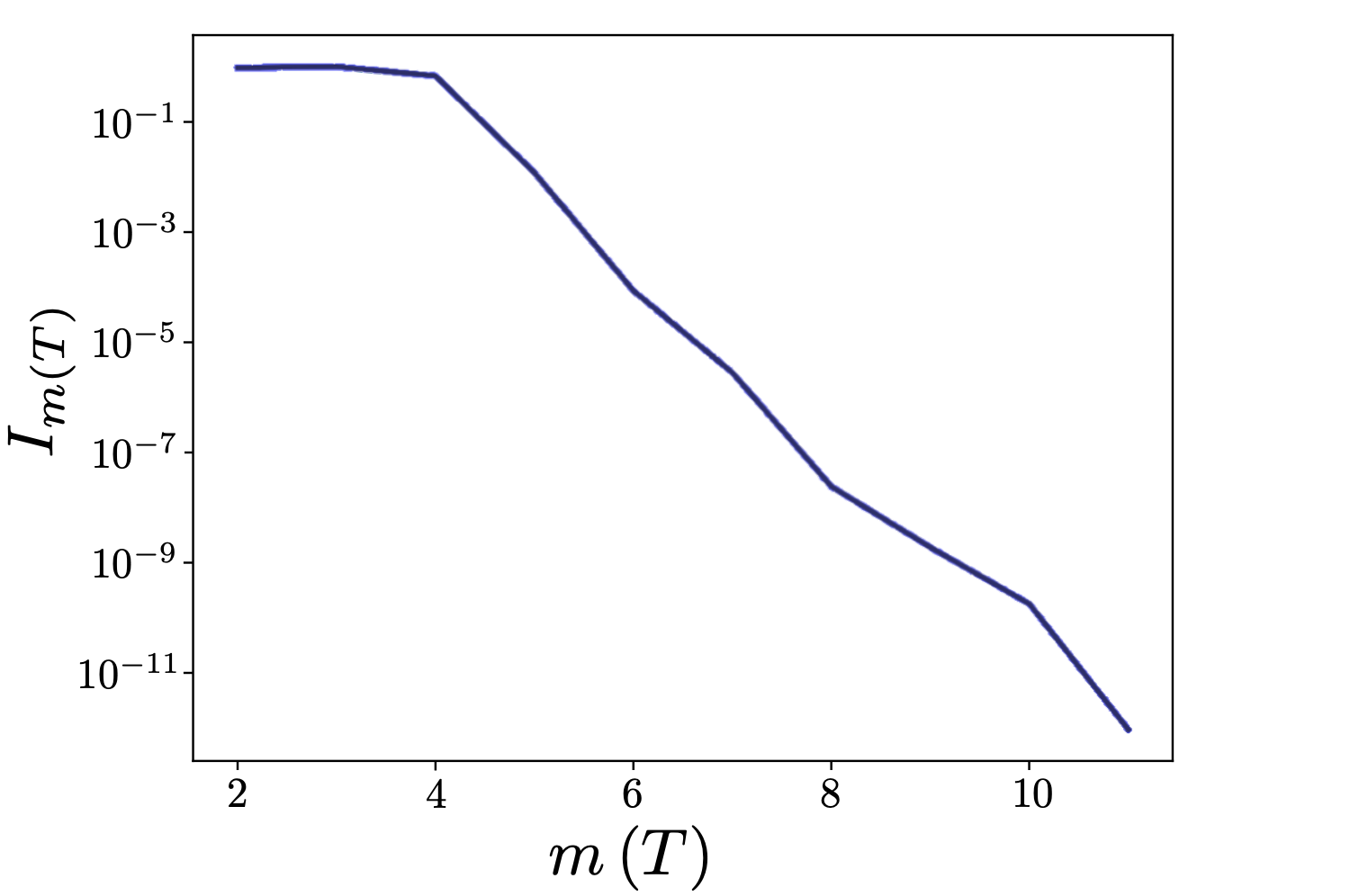}
\caption{\label{fig:Square-root-infidelity-min} Square-root infidelity between the state $\left|\Psi\left(T\right)\right\rangle ^{\prime}$ obtained for $\hat{H}_{+ \rm min}\left(t\right)$ eq. (\ref{eq:H_min}), and the state $\left|\Psi\left(T\right)\right\rangle$ containing all
modes $\hat{\kappa}_{k}^{\rm in}$. The nonstationary Schrodinger equation was solved numerically in a Fock space which was truncated at a maximal number of environment's quanta $n_{\rm cut}=4$ for $T=100$. The driven qubit $\hat{H}_{s}\left(t\right)=\hat{\sigma}_{+}\hat{\sigma}_{-}+\hat{\sigma}_{x}0.1\cos t$
was taken as an open system, $g=0.1$, $\hat{V}_{s}=\hat{\sigma}_{-}$,
$\varepsilon_{j}\equiv1$, $h_{j}\equiv0.05$. The semiinfinite chain eq. (\ref{eq:semiinf_chain}) was truncated at $30$ sites.}
\end{figure}

In Fig. \ref{fig:Square-root-infidelity-min} we show the dependence of the infidelity $I_{m(T)}$ on $m(T)$ between $\ket{\Psi\left(T\right)}^{+ \rm min}$ evolved with effective Hamiltonian of minimal light cone $\hat{H}_{+ \rm min}\left(t\right)$ and the state $\ket{\Psi\left(T\right)}$ for which all the modes $\hat{\kappa}_{k}^{\rm in \dagger}$ are kept (without sum truncation).

\subsection{Minimal backward light cone with stable records}

In the previous section, we identified the environmental degrees of freedom that significantly interacted with the open system. To ensure that these modes can store stable records and enable the construction of projections for decoherent histories, we must define the irreversibly decoupled modes (iv) --- those that no longer interact significantly with the system.

In the same way we use a future-averaged OTOC:

\begin{equation}
\int\limits_{t}^{T} d\tau \,\,\mathcal{C}\left(\kappa,\tau\right) = \left\langle \kappa\right|\hat{\rho}_{-}\left(t\right)\left|\kappa\right\rangle
\label{eq:future_OTOC}
\end{equation}
where $\hat{\rho}_{-}\left(t\right)=\int\limits_{t}^{T}d\tau\left|\alpha\left(\tau\right)\right\rangle \left\langle \alpha\left(\tau\right)\right|\label{eq:rho_minus}$ with $\alpha\left(\tau\right)$ being the solution of eqs. (\ref{eq:a0_spread_eq}-\ref{eq:a0_spread_eq_ini_cond}).

If this quantity falls below a threshold, the mode is considered decoupled. Two types of decoupled modes can be distinguished. The first type consists of modes that never interacted with the open system; they carry no information and can be discarded. The second type consists of modes that interacted in the past but have since irreversibly decoupled; these modes store records of the system’s history.

To isolate these record-carrying modes, we search for irreversibly decoupled modes within the subspace of previously coupled ones $\kappa_{1}^{\rm in}\ldots\kappa_{m_{+}\left(t\right)}^{\rm in}$, using the absence of significant future interaction as a criterion:
\begin{equation}
g_{-}\left(\chi,t\right)=\left\langle \chi\right|\rho_{-}^{m_{+}\left(t\right)}\left(t\right)\left|\chi\right\rangle -a_{\rm cut} < 0,
\label{eq:backwardLC}
\end{equation}
where $\hat{\chi}^{\dagger}=\sum\limits_{p=1}^{m_{+}\left(t\right)}\chi_{p}\hat{\kappa}_{p}^{\rm in \dagger}$, $\ket{\chi} $ denotes the vector with components
$\chi_{p}$
and 
\begin{equation}
\left[\rho_{-}^{m_{+}\left(t\right)}\left(t\right)\right]_{pq}=\left\langle \kappa_{p}^{\rm in}\left|\hat{\rho}_{-}\left(t\right)\right|\kappa_{q}^{\rm in}\right\rangle 
\label{eq:rho_minus_proj}
\end{equation}
for $p,q = 1,\ldots, m_{+}\left(t\right)$. The contribution of such modes to the system’s future dynamics becomes negligible after time $t$.

We define the escape time $t_{\rm out}(\kappa)$ as the latest moment at which mode $\kappa$ contributes significantly to the system’s evolution:
\begin{equation}
t_{\rm out}\left(\chi\right)=\int_{0}^{T}d\tau\,\theta\left[g_{-}\left(\chi,\tau\right)\right]\label{eq:escape_time}
\end{equation}

The minimal backward light cone identifies the environmental modes that have irreversibly decoupled from the system $\kappa_{1}^{\rm out}\ldots\kappa_{m_{-}\left(T\right)}^{\rm out}$ and thus carry stable records. The number of modes which have escaped before $t$:
\begin{equation}
m_-\left(t\right)=\intop_{0}^{t}d\tau\sum_{p=1}^{m\left(T\right)}\delta\left(\tau-t_{\rm out}\left(\kappa_{p}^{\rm out}\right)\right).
\label{eq:m_minus}
\end{equation}

These modes provide the natural subspace in which to define the projectors for decoherent histories. Details of the construction are given in \cite{Pol2022}.

\begin{figure}
\centering
\includegraphics[width=0.48\textwidth]{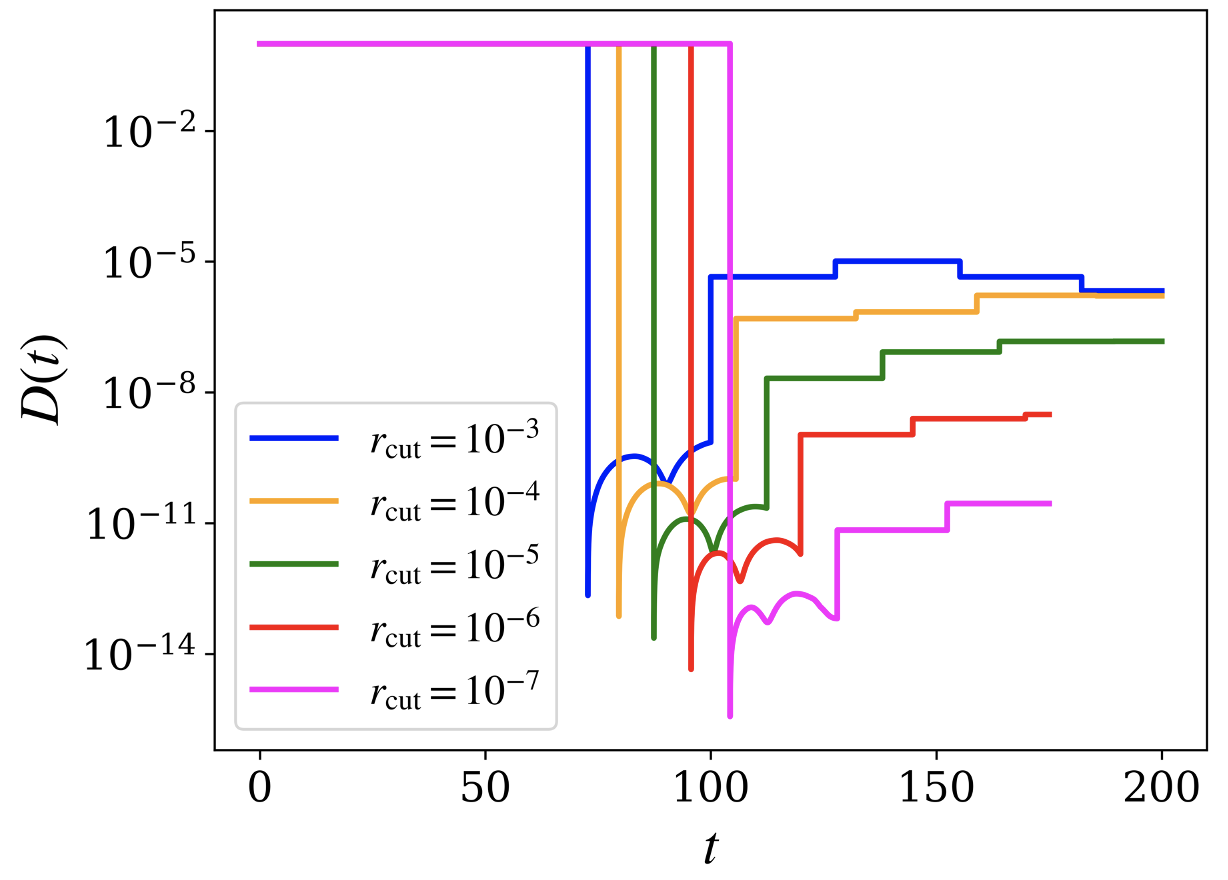}
\caption{
The plot shows the average decoherence overlap (\ref{eq:DO}) over 50 sampled histories as a function of time, for different values of the significance threshold $r_{\rm cut}$. The number of modes inside minimal light cone computed with different significance threshold $r_{\rm cut}$. The Fock space was truncated at a maximal number quanta $n_{\rm cut} = 5$ for $T=150$ and $T=200$. The driven qubit $\hat{H}_{s}\left(t\right)=\hat{\sigma}_{+}\hat{\sigma}_{-}+\hat{\sigma}_{x}\,0.1\,\cos t$
was taken as an open system, $g=0.05$, $\hat{V}_{s}=\hat{\sigma}_{-}$,
$\varepsilon_{j}\equiv1$, $h_{j}\equiv0.05$. Here no truncation to the relevant modes was applied; otherwise the overlaps would vanish exactly after each projection.}
\label{fig:DC_time}
\end{figure}

\section{\label{sec:dec_over} Decoherence overlap}
By extracting the irreversibly decoupled modes $\kappa_{1}^{\rm out}\ldots\kappa_{m_{-}\left(T\right)}^{\rm out}$, we construct a decoherent history emerging at the successive escape times $t_{\rm out}(\kappa_{1}^{\rm out})\ldots t_{\rm out}(\kappa_{m_{-}\left(T\right)}^{\rm out})$.

The coupling of the irreversibly decoupled modes to the future quench dynamics is below the statistical significance threshold $r_{\rm cut}$. 
The system evolution on $\left[0,t\right]$ can therefore be effectively reduced to the relevant modes inside the minimal forward light cone that have not yet irreversibly decoupled (iii). Denoting their number by $r(t) = m_+(t) - m_-(t)$, the effective Hamiltonian in the moving basis takes the form:
\begin{equation}
\begin{split}
\hat{H}\left(t\right) \approx \hat{H}_{\rm eff}(t) = \hat{H}_s + g\,\hat{V}_{s}^{\dagger}\sum_{k=1}^{r(t)} \chi_{k}^{*}\left(t\right)\hat{\kappa}_{k}^{\rm rel}\,+\\+\,g\,\hat{V}_{s}\sum_{k=1}^{r(t)}\chi_{k}\left(t\right)\hat{\kappa}_{k}^{\rm rel}{}^\dagger - \sum_{k=1}^{r(t)}\sum_{l=1}^{r(t)}\xi_{kl}\left(t\right)\hat{\kappa}_{k}^{\rm rel}{}^\dagger\hat{\kappa}_{l}^{\rm rel}
\end{split}
\label{eq:eff_Ham}
\end{equation}
where $\chi_{k}(t)$ are the time-dependent coupling amplitudes,  $\xi_{kl}(t)$ gradually implement the rotation and together with $\hat{\kappa}_k$ are defined by the non-stationary Bogoliubov transformation (see Appendix~\ref{app:trans_moving_basis}, \cite{Pol2022}).

Consequently, the projections onto eigensubspaces of irreversibly decoupled modes commute with the interaction Hamiltonian after escape time:
\begin{equation}
\left[\hat{H}_{\rm int}(t), \hat{P}_{{\kappa^{\rm out}};\, n_{\kappa^{\rm out}}}\right] = O\left(\sqrt{r_{\rm cut}}\right) \approx 0
\end{equation}
for $t > t_{\rm out}(\kappa^{\rm out})$ up to an accuracy controlled by significance threshold $r_{\rm cut}$ (see Appendix~\ref{app:err_bound}). For the projectors we use occupation number eigensubspace: 
\begin{equation}
\hat{n}_{p}^{\rm out}\hat{P}_{{\kappa_p^{\rm out}};\, n_p^{\rm out}} = n_p^{\rm out}\hat{P}_{{\kappa_p^{\rm out}};\, n_p^{\rm out}}  
\label{eq:eigenspace}
\end{equation} so that each irreversibly decoupled mode $\kappa^{\rm out}$ carries a well-defined and stable record about system past. 

\begin{figure}[!htb]
\includegraphics[width=0.45\textwidth]{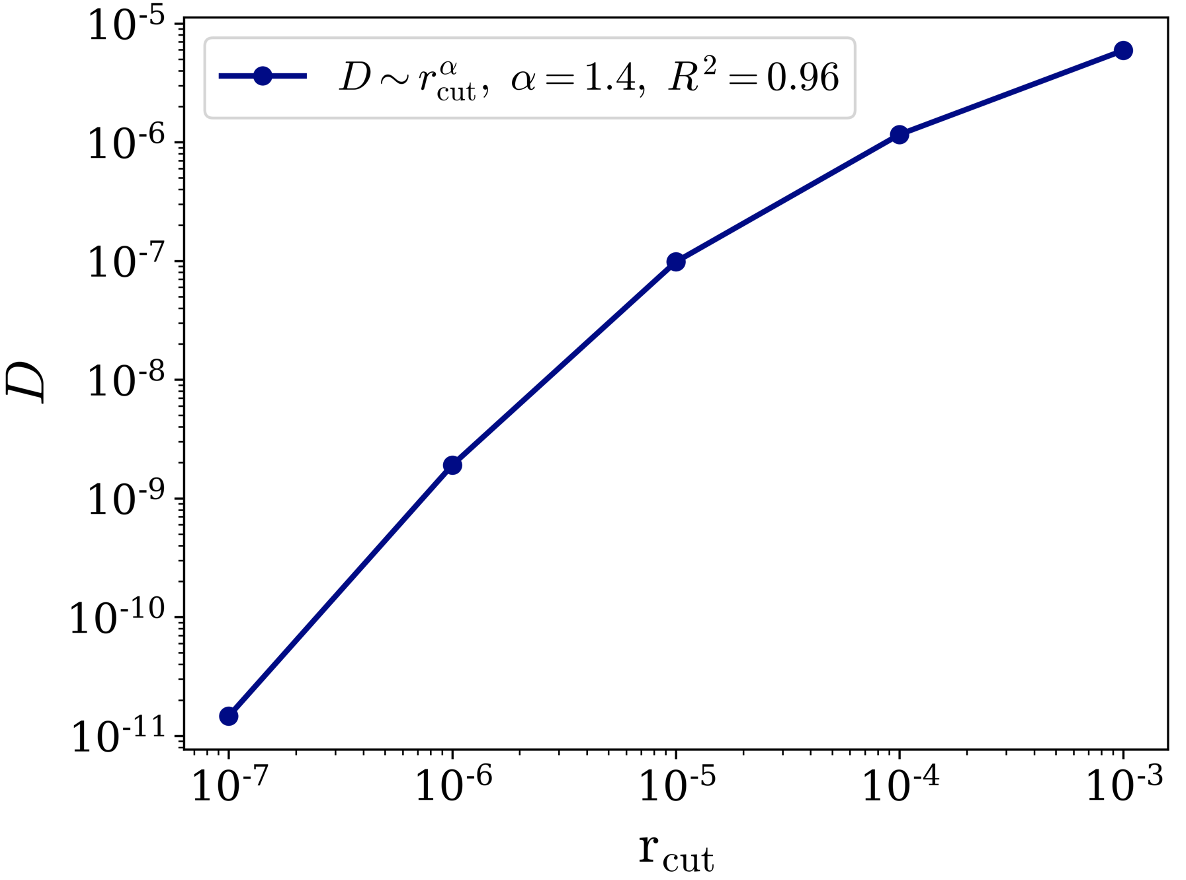}
\caption{
The time dependence of the mean plateau value of the decoherence functional. Here no truncation to the relevant modes was applied; otherwise the overlaps would vanish exactly after each projection.}
\label{fig:mean_DO}
\end{figure}

The corresponding history state is:
\begin{equation}
\begin{split}
\ket{\Psi_{\boldsymbol{\alpha}}(t)} = \hat{P}_{\kappa_{m_-(t)}^{\rm out};\,n_{m_-(t)}^{\rm out}}\hat{U}(t_{\kappa_{m_-(t)}^{\rm out}}, t_{\kappa_{m_-(t)-1}^{\rm out}})\ldots
\\
\ldots \hat{U}(t_{\kappa_2^{\rm out}},t_{\kappa_1^{\rm out}})\hat{P}_{\kappa_{1}^{\rm out};\,n_1^{\rm out}}\hat{U}(t_{\kappa_1^{\rm out}})\ket{\Psi(0)}
\end{split}\label{eq:hist_modes}
\end{equation}
with pure initial condition (\ref{eq:initial_cond}). Here $\hat{U}(t_i,t_j)$ is full time evolution.

The decoherence overlap between two distinct histories $\boldsymbol{\alpha} = (n_1^{\rm out}, ... n_{m_-(t)}^{\rm out})$ is: 
\begin{equation}
\mathcal{D}(\boldsymbol{\alpha},\boldsymbol{\beta})(t)=\bra{\Psi_{\boldsymbol{\beta}}(t)}\left.\Psi_{\boldsymbol{\alpha}}(t)\right\rangle \approx 0\,,\,\, \text{for  } \boldsymbol{\alpha}\ne\boldsymbol{\beta}\label{eq:DO_modes}
\end{equation}
up to the accuracy controlled by the significance threshold, demonstrating approximate diagonality (see Appendix~\ref{app:err_bound}).

Although the tape-recorder coarse-graining drastically reduces the Hilbert space by retaining only modes inside the minimal light cone, the number of possible histories still grows as \(N_{\rm hist}=n^k\), where \(n\) is the number of alternatives at each coarse-graining step $k$. The number of nontrivial overlaps is then \(n^{2k}-n^k-l\), since there are \(n^k\) diagonal terms equal one, and \(l = n(n-1)n^{2(k-1)}\) trivial zero overlaps due to the final-time $t_k$ mismatch \(n^{\rm out}_k\neq n^{\prime\, \rm out}_k\). In practice, we evaluate averages over a subset of histories. 

The Fig.~\ref{fig:DC_time} illustrates the average decoherence overlap (\ref{eq:DO}) time dependence for different significance thresholds $r_{\rm cut}$ computed over 50 sampled histories. 
The Fig.~\ref{fig:mean_DO} shows the average decoherence overlap plateau depending on $r_{\rm cut}$, computed over 50 sampled histories. These results confirm that the identified modes are indeed irreversibly decoupled and that projections onto their subspaces yield decoherent histories.

Importantly, irreversibly decoupled modes appear at characteristic times that remain essentially unchanged upon increasing the total evolution time $T$ (see Appendix~\ref{app:trans_moving_basis}, Figs.~\ref{fig:t_out_vs_T},\ref{fig:TT}).This ensures that the decoherence overlap remains stable for finite-time simulations.

\subsection{Physical meaning of significance threshold $r_{\rm cut}$}
As demonstrated by the infidelity of many-body joint state of system and environment $I(t)$ (\ref{eq:infidelity}, \ref{eq:rel_guard_modes_convergence}), the contribution of the discarded modes to the dynamics is proportional to $r_{\rm cut}$, and can be made arbitrarily small by decreasing the threshold, ensuring that observables remain accurately reproduced.

\begin{figure}
\centering
\includegraphics[width=0.49\textwidth]{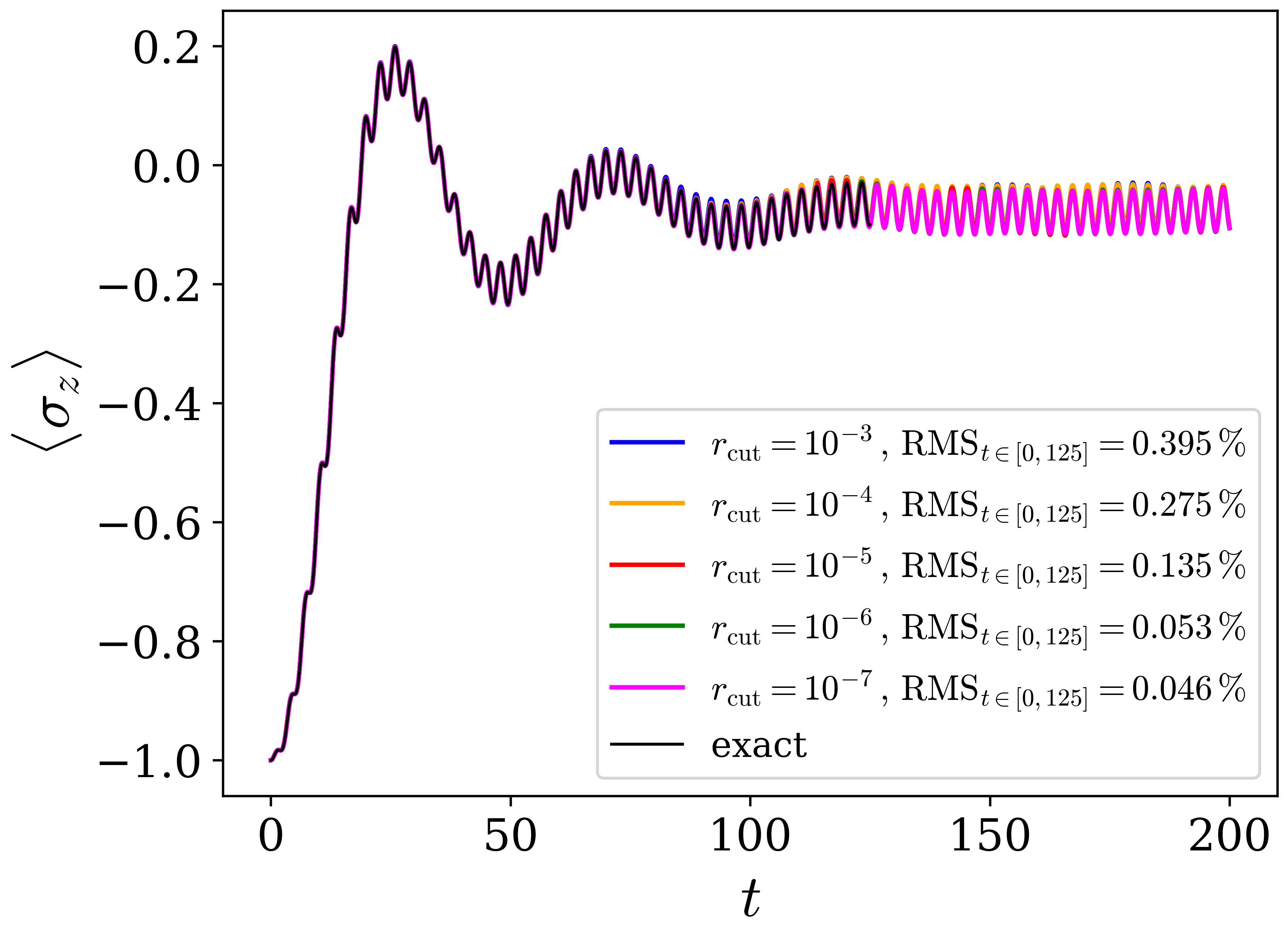}
\caption{Time evolution of $\langle\sigma_z\rangle$. The irreversibly decoupled modes are traced out using the quantum jump Monte Carlo simulation, providing an efficient unraveling of the full wave function into a sampled ensemble of 300 decoherent histories. The value of the  root mean square deviation (RMS) is shown in the figure corresponding to the range of times from 0 to 125 and decrease with $r_{\rm cut}$.}
\label{fig:sigma_z}
\end{figure}

Conceptually, $r_{\rm cut}$ has an operational meaning: it is determined by the smallest coherences that can be resolved in an experiment. Any measurement setup has finite time resolution, bandwidth, and signal-to-noise ratio, while environmental effects such as thermal fluctuations further limit coherence detection \cite{Clerk2010}, setting a lower bound on the experimentally achievable infidelity. Residual coherences between different decoherent histories then lie below this resolution and are operationally unobservable. This defines a physically meaningful distinguishability scale, which naturally identifies the appropriate range of $r_{\rm cut}$.

Concretely, $r_{\rm cut}$ should be chosen from infidelity bound (\ref{eq:infidelity}):
\begin{equation}
r_{\rm cut} \propto a_{\rm cut} = \dfrac{I_{\rm experimental}}{\left(g\Vert \hat{V}_s\Vert\right)^2} 
\end{equation}
where $I_{\rm experimental}$ is the experimentally achievable infidelity.

\begin{figure}
\includegraphics[width=0.4\textwidth]
{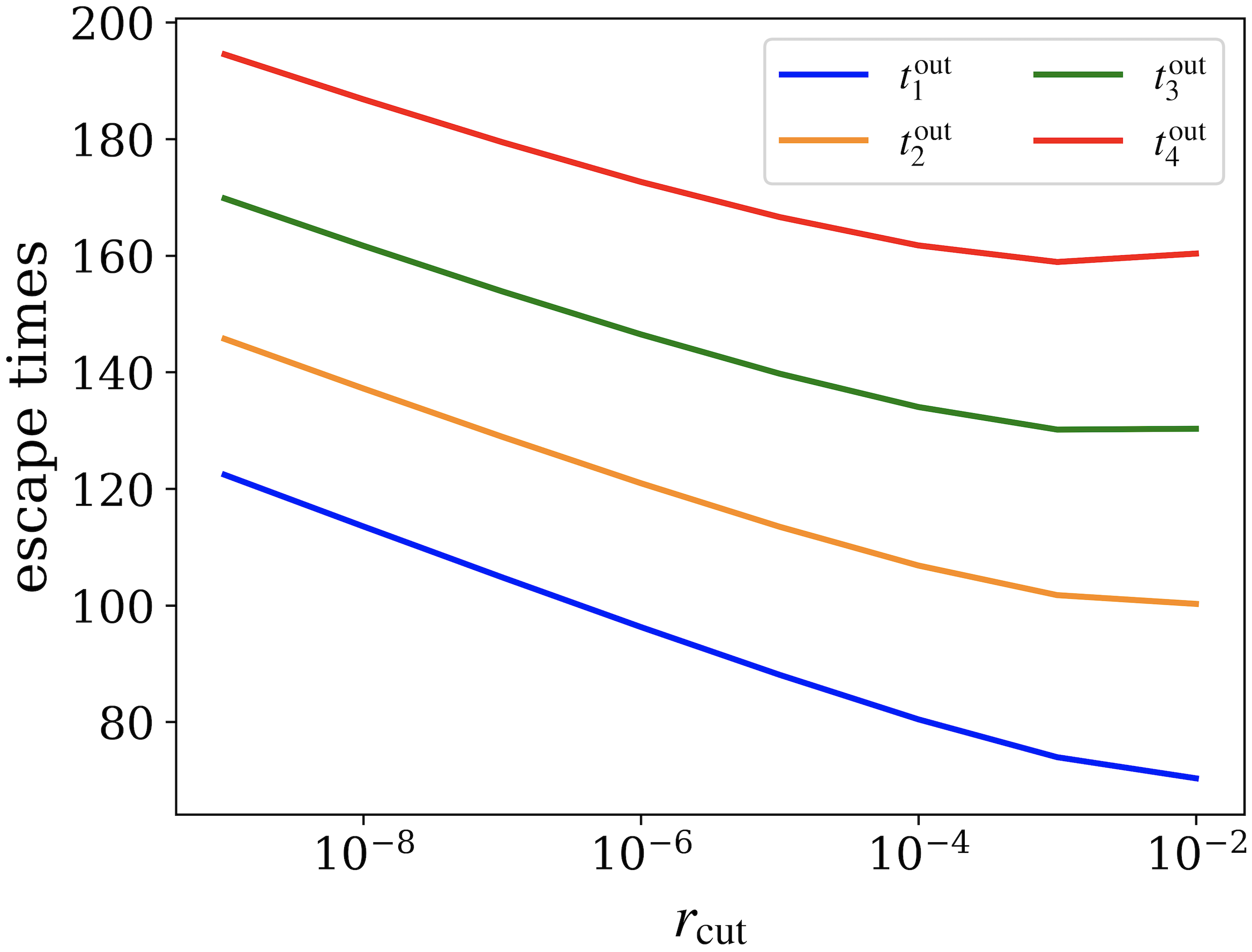}
\caption{The dependence of escape time $t^{\rm out}$ on $r_{\rm cut}$.}
\label{fig:times_out}
\end{figure}

Importantly, the precise numerical value of $r_{\rm cut}$ is not critical. Observables are robust as shown in Fig.~\ref{fig:sigma_z}, and the escape time $t^{\rm out}$ depends only logarithmically on $r_{\rm cut}$ (see Fig.~\ref{fig:times_out}), so variations of the threshold within the experimentally admissible range produce only minor changes. This demonstrates that the emergence of classical records is insensitive to the exact choice of $r_{\rm cut}$, reflecting the objective and operational nature of the classical limit.

\section{\label{sec:unrav} Unraveling of dynamics}

A particularly appealing consequence of the decoherent histories framework is that the quantum evolution admits an approximate stochastic unraveling in the space of decoherent histories. In this sense, classical reality is encoded in the ensemble of emergent decoherent histories.

The total state of the system corresponds to the sum over all histories:
\begin{equation}
\ket{\Psi(t)} = \sum_{\boldsymbol{\alpha}} \ket{\Psi_{\boldsymbol{\alpha}}(t)} ,
\end{equation}

However, the number of such histories grows exponentially with time and quickly becomes computationally intractable. To obtain an efficient description, we replace the full state by a stochastic sampling of individual histories. Once a mode becomes irreversibly decoupled from the system, meaning its influence on future evolution falls below a significance threshold $r_{\rm cut}$, its degrees of freedom can be traced out. Using (\ref{eq:hist_modes}) this corresponds to evolving with the effective Hamiltonian only with relevant modes (\ref{eq:eff_Ham}):
\begin{equation}
\hat U_{\rm eff}(t_i,t_j)\approx
\mathcal T
\exp\!\left[
-i\int_{t_j}^{t_i} d\tau\, \hat H_{\rm eff}(\tau)
\right].
\end{equation}
This leads to an effective reduced density matrix:
\begin{equation}
\rho(t) \approx \mathrm{Tr}_{\kappa_1^{\mathrm{out}},\dots,\kappa_{m_-(t)}^{\mathrm{out}}}
\ket{\Psi(t)}\bra{\Psi(t)} .
\end{equation}

Operationally, this procedure is analogous to a quantum jump: each time a mode becomes irreversibly decoupled the state is projected onto one of the corresponding branches and renormalized. In practice, we implement this by Monte-Carlo sampling. 

Figure~\ref{fig:sigma_z} was obtained in this way, by averaging over 300 randomly generated histories for different significant threshold $r_{\rm cut}$, defining the number of relevant modes.

\section{\label{sec:discus} Discussion}
We have shown that the tape-recorder coarse-graining provides a physically grounded way to construct decoherent histories in integrable, non-Markovian environments. Although our explicit analysis was performed for a spin-boson model, the underlying mechanism is quite general and we expect that it can be extended to a broad class of open quantum systems, including those with chaotic environments. 

Current research on out-of-equilibrium many-body quantum dynamics focuses on questions such as how quenched systems approach equilibrium; how the correlations and quantum information propagate throughout the system; how factors such as integrability, disorder affect the quench dynamics \cite{Eisert2015}. Concepts such as operator growth, entanglement spreading, and out-of-time-order correlators (OTOCs) \cite{Xu:2022,Hashimoto:2017} play a central role in these studies. Our approach complements it and provides a framework for explicitly constructing multi-time decoherent histories, in which the system evolves along a branching structure of mutually exclusive alternatives.

The key carriers of information in this construction are the irreversibly decoupled environmental modes, which act as stable records of the system's past. Each history corresponds to an effective sequence of quantum jumps, so that the full many-body dynamics can be interpreted as an unraveling into a branching ensemble of trajectories. This enables efficient Monte Carlo sampling of histories, as illustrated by the agreement between observables computed via exact dynamics and those reconstructed from history unravelings (Fig.~\ref{fig:sigma_z}).

The significance threshold $r_{\rm cut}$, which determines the coarse-graining of the environment, is not an artificial parameter, it plays a role analogous to the noise floor of the recording head in the tape recorder picture. Its physically motivated and practically optimal choice is the experimentally achievable infidelity. Just as a tape recorder cannot meaningfully distinguish a signal below its noise floor, quantum superpositions with residual coherences below the infidelity are operationally indistinguishable from classical mixtures.

With this choice, the off-diagonal elements of the decoherence functional --- the interference terms between distinct histories --- are suppressed to order $O(\sqrt{r_{\rm cut}}) = O(\sqrt{I})$, making the histories decoherent at the level of operational distinguishability. As a result, the construction is uniform in time and does not exhibit a breakdown time within the considered dynamics.

Beyond the explicit construction of histories, the tape-recorder coarse-graining naturally opens new perspectives. In particular, in earlier work we defined the entropy of decoherent histories and argued that it can serve as a diagnostic tool of quantum chaos \cite{PolAref:24}. Establishing a systematic connection to non-equilibrium thermodynamics remains an open question for future study.

Finally, we note that our approach appears amenable to experimental tests. Platforms such as trapped ions, superconducting qubits, and cold atoms already allow for local quenches and partial access to environmental degrees of freedom. These capabilities may enable direct probing of irreversibly decoupled modes and the associated entropy of decoherent histories.

\section{\label{sec:concl}Conclusions}
We introduced a mechanism for the emergence of decoherent histories in integrable, non-Markovian environments, based on viewing the environment as a tape recorder that sequentially and irreversibly stores information about the system's past. This provides a controllable and explicit way to construct consistent multi-time histories without assuming non-integrability or Markovianity. Decoherence is controlled by a physically meaningful threshold $r_{\rm cut}$ set by the experimentally accessible infidelity, and the resulting suppression of interference remains stable in time, with residual interference bounded by the infidelity. This makes the approach practically applicable beyond the short-time regime.

Our results suggest that the decoherent histories framework can be developed into a practical tool for studying out-of-equilibrium quantum dynamics. At the same time, they point to open directions, such as extending the analysis to more complex environments and clarifying the role of history entropy. In this sense, the tape-recorder coarse-graining complements Quantum Darwinism by emphasizing the sequential and multi-time character of decoherent histories.

\subsection*{Acknowledgement}
The work of N. A. was supported by the Theoretical Physics and Mathematics Advancement Foundation “BASIS” Grant No. 23-2-2-26-1.

\newpage

\appendix
\section{\label{app:just_Lieb_Rob} Infidelity bound and justification of the Lieb-Robinson metric}

To establish the convergence of the method, we demonstrate that modes outside the minimal light cone can be neglected in the system's evolution.

We consider the evolution eq. (\ref{eq:initial_cond}-\ref{eq:interaction_pic_sch_eq}) on the interval $\left[0,t\right]$. The system-environment coupling can be decomposed into contributions from modes inside and outside the light cone as:
\begin{equation}
\hat{a}_{0}^{\dagger}\left(t\right)=\hat{a}_{\perp}^{\dagger}\left(t\right)+\eta\left(t\right)\hat{\kappa}^\dagger{}^{\rm edge},
\end{equation}
where $\left[\hat{a}_{\perp}\left(t\right),\hat{\kappa}^{\dagger}{}^{\rm edge}\right]=0$
and $\eta\left(t\right)=\left[\hat{\kappa}^{\rm edge},\hat{a}_{0}^{\dagger}\left(t\right)\right]_{\pm}=\left\langle \kappa^{\rm edge}\left|\alpha\left(t\right)\right.\right\rangle$. 

\begin{figure}
\includegraphics[width=0.5\textwidth]{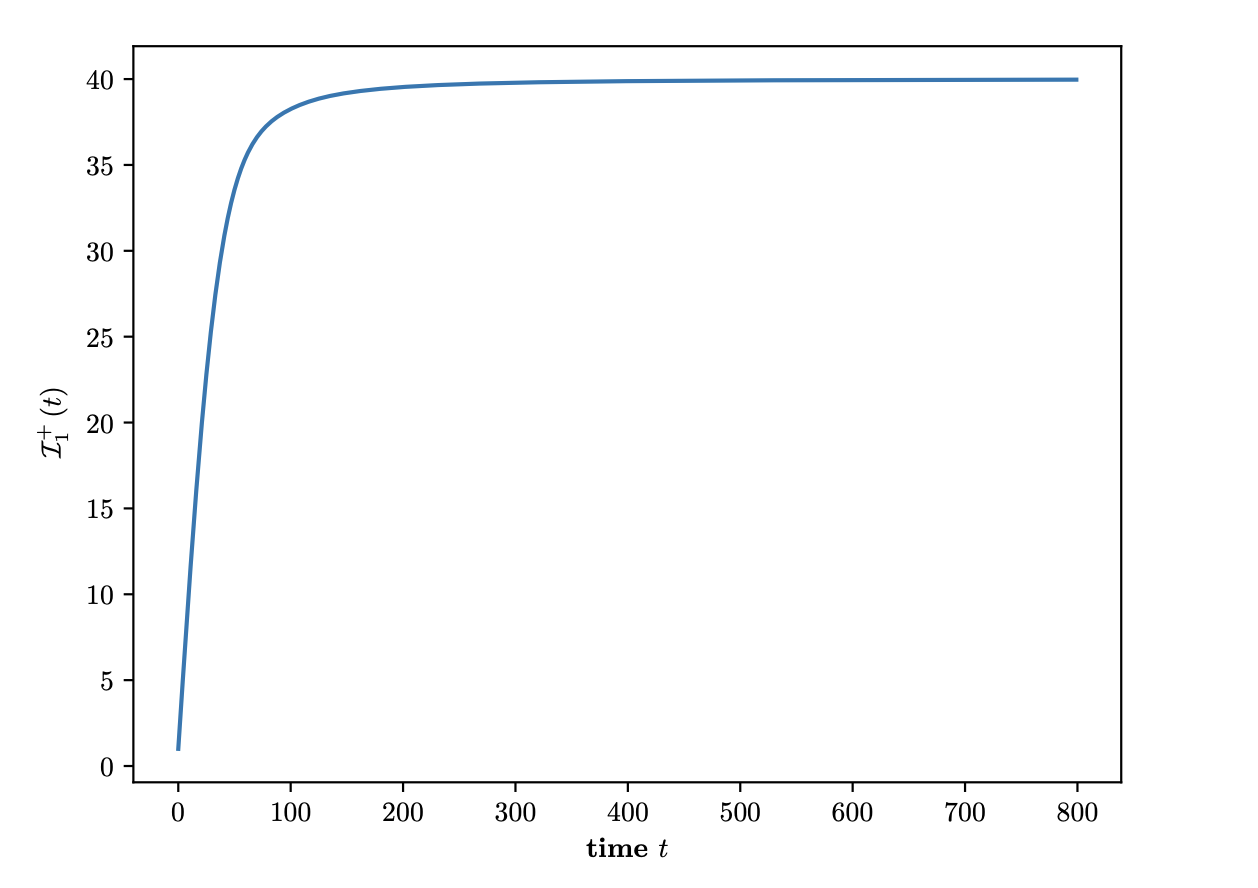}
\caption{\label{fig:Plot-of-the_I1_plus_t}Plot of the largest eigenvalue $\mathcal{I}_{1}^{+}\left(t\right)$ of the retarded light cone density matrix $\hat{\rho}_{+}\left(t\right)$. It is seen that at large times $\mathcal{I}_{1}^{+}\left(t\right)$ saturates at some finite value $\mathcal{I}_{1}^{+}\left(\infty\right)$. }
\end{figure}

We introduce the Hamiltonian with the $\hat{\kappa}^{\rm edge}, \hat{\kappa}^{\rm edge\dagger}$ modes discarded:
\begin{equation}
\hat{H}_{\perp}\left(t\right)=\hat{H}_{s}+g\hat{V}_{s}^{\dagger}\hat{a}_{\perp}\left(t\right)+g\hat{V}_{s}\hat{a}_{\perp}^{\dagger}\left(t\right).\label{eq:oqs_interaction_pic-1-1}
\end{equation}
Suppose $\left|\Psi_{\perp}\left(t\right)\right\rangle $ is the evolution under $\hat{H}_{\perp}\left(t\right)$:
\begin{equation}
\left|\Psi_{\perp}\left(t\right)\right\rangle =\mathcal{T}\exp\left[-i\intop_{0}^{t}d\tau\hat{H}_{\perp}\left(\tau\right)\right]\left|\Psi\left(0\right)\right\rangle
\end{equation}
where $\mathcal{T}$ is the time ordering operator. To quantify the effect of neglecting the edge modes, we define the square-root fidelity:
\begin{equation}
\sqrt{F(t)}=\left|
\langle\Psi_{\perp}(t)|\Psi(t)\rangle\right|
\label{eq:sqrt_fidelity1}
\end{equation}
The corresponding infidelity: 
\begin{equation}
I\left(t\right)=1-\sqrt{F\left(t\right)}\label{eq:infidelity1}
\end{equation}
provides a measure of the statistical significance of the mode $\kappa^{\rm edge}$ over the interval $[0,t]$. For brevity, we denote the edge mode $\kappa^{\rm edge}$ simply by $\kappa$.

Expanding the infidelity in powers of $\kappa$ gives: 
\begin{equation}
I\left(\kappa,t\right)=I_{0}\left(\kappa,t\right)+I_{2}\left(\kappa,t\right)+\ldots+I_{2p}\left(\kappa,t\right)+\ldots.\label{eq:infidelity_expansion}
\end{equation}
Only even-order terms appear in this expansion because the vacuum expectation values of odd powers of $\hat{\kappa}, \hat{\kappa}^{\dagger}$ vanish in eq.~(\ref{eq:sqrt_fidelity1}). Therefore, each $\hat{\kappa}^{\dagger}$ should be annihilated by some $\hat{\kappa}$, which contributes factor $\eta^{*}\left(\tau^{\prime}\right)\eta\left(\tau\right)$. At order zero we have $\Psi\left(t\right)=\Psi_{\perp}\left(t\right)$, therefore $I_{0}\left(\kappa,t\right)=0$. The term $I_{2p}\left(\kappa,t\right)$ can be written as:
\begin{widetext}
\begin{equation}
I_{2p}\left(\kappa,t\right)
=\left[\frac{1}{p!}\right]^{2}\left\langle \Psi_{\perp}\left(t\right)\right|\intop_{0}^{t}d\tau_{1}\ldots d\tau_{p}d\tau_{1}^{\prime}\ldots d\tau_{p}^{\prime}
\mathcal{T}\prod_{i=1}^{p}\left[g\eta^{*}\left(\tau_{i}\right)\hat{V}_{s}^{\dagger}\hat{\kappa}\right]\prod_{j=1}^{p}\left[g\eta\left(\tau_{j}^{\prime}\right)\hat{V}_{s}\hat{\kappa}^{\dagger}\right]
\exp\left[-i\intop_{0}^{t}d\tau\hat{H}_{\perp}\left(\tau\right)\right]\left|\Psi\left(0\right)\right\rangle .\label{eq:_I_2p}
\end{equation}
\end{widetext}
Here we first apply time ordering to the operators to the right of $\mathcal{T}$ (including the operator exponential). Then we commute all $\hat{\kappa}$ to the right. This produces some additional factor $\mathcal{C}\leq p!$. Therefore, this expression can be bounded from above:
\begin{equation}
\left|I_{2p}\left(\kappa, t\right)\right|\leq\left[\frac{1}{p!}\right]^{2}p!\left[g\left\Vert \hat{V}_{s}\right\Vert \right]^{2p}\left\Vert \eta\right\Vert _{L^{2}}^{2p},
\end{equation}
where 
\begin{equation}
\left\Vert \eta\right\Vert _{L^{2}}^{2}=\intop_{0}^{t}d\tau\left|\eta\left(\tau\right)\right|^{2}=\intop_{0}^{t}d\tau\left|\left\langle \kappa\left|\alpha\left(\tau\right)\right.\right\rangle \right|^{2}=\left\langle \kappa\right|\hat{\rho}_{+}\left(t\right)\left|\kappa\right\rangle ,
\end{equation}
and we obtain:
\begin{equation}
\left|I_{2p}\left( \kappa, t\right)\right|\leq\frac{1}{p!}\left\{ \left[g\left\Vert \hat{V}_{s}\right\Vert \right]^{2}\left\langle \kappa\right|\hat{\rho}_{+}\left(t\right)\left|\kappa\right\rangle \right\} ^{p}.
\end{equation}

The infidelity expansion (\ref{eq:infidelity_expansion}) is bounded from above as:
\begin{widetext}
\begin{equation}
I\left(\kappa,t\right)\leq\left|I_{2}\left(\kappa,t\right)\right|+\ldots+\left|I_{2p}\left(\kappa,t\right)\right|+\ldots=\sum_{p=1}^{\infty}\frac{1}{p!}\left\{ \left[g\left\Vert \hat{V}_{s}\right\Vert \right]^{2}\left\langle \kappa\right|\hat{\rho}_{+}\left(t\right)\left|\kappa\right\rangle \right\} ^{p}
=\exp\left(\left[g\left\Vert \hat{V}_{s}\right\Vert \right]^{2}\left\langle \kappa\right|\hat{\rho}_{+}\left(t\right)\left|\kappa\right\rangle \right)-1
.\label{eq:infidelity_bound}
\end{equation}
\end{widetext}

From the light cone interior condition (\ref{eq:Lieb_robinson_metric}) we obtain that if the modes outside the light cone are neglected, the resulting infidelity is:
\begin{equation}
I\left( t\right)\lesssim\left[g\left\Vert \hat{V}_{s}\right\Vert \right]^{2}a_{\rm cut},\label{eq:convergence_with_respect_to_a_cut}
\end{equation}
that is the full many body wavefunction converges as $a_{\rm cut}\to0$.

Observe the important property that the final time $t$ does not enter the convergence estimate (\ref{eq:convergence_with_respect_to_a_cut}). That is, we obtain some approximation which is converging uniformly in time.

It appears that the convergence is of first order in regularization $a_{\rm cut}$. However, describing convergence in terms of $a_{\rm cut}$ is not physically meaningful. Actually, when $a_{\rm cut}\to0$, the number of independent modes $m\left(t\right)$ which are inside the lightcone is increasing. A more physically relevant characterization is the convergence of the infidelity as a function of the number of retained modes $m\left(t\right)$.

For parameters $\varepsilon_{j}\equiv1$, $h_{j}\equiv0.05$ of the semiinfinite chain (\ref{eq:semiinf_chain}) and varying $t=0\ldots800$, we find $\mathcal{I}_{p}^{+}\left(t\right)$ and $\left|\widetilde{\kappa}_{p}\right\rangle $ (\ref{eq:rho_plus_eigvals}). In Fig.~\ref{fig:Plot-of-the_I1_plus_t} we plot $\mathcal{I}_{1}^{+}\left(t\right)$. It saturates at some finite value $\mathcal{I}_{1}^{+}\left(\infty\right)$ for large times. 

\begin{figure}[!htb]
\includegraphics[width=0.52\textwidth]{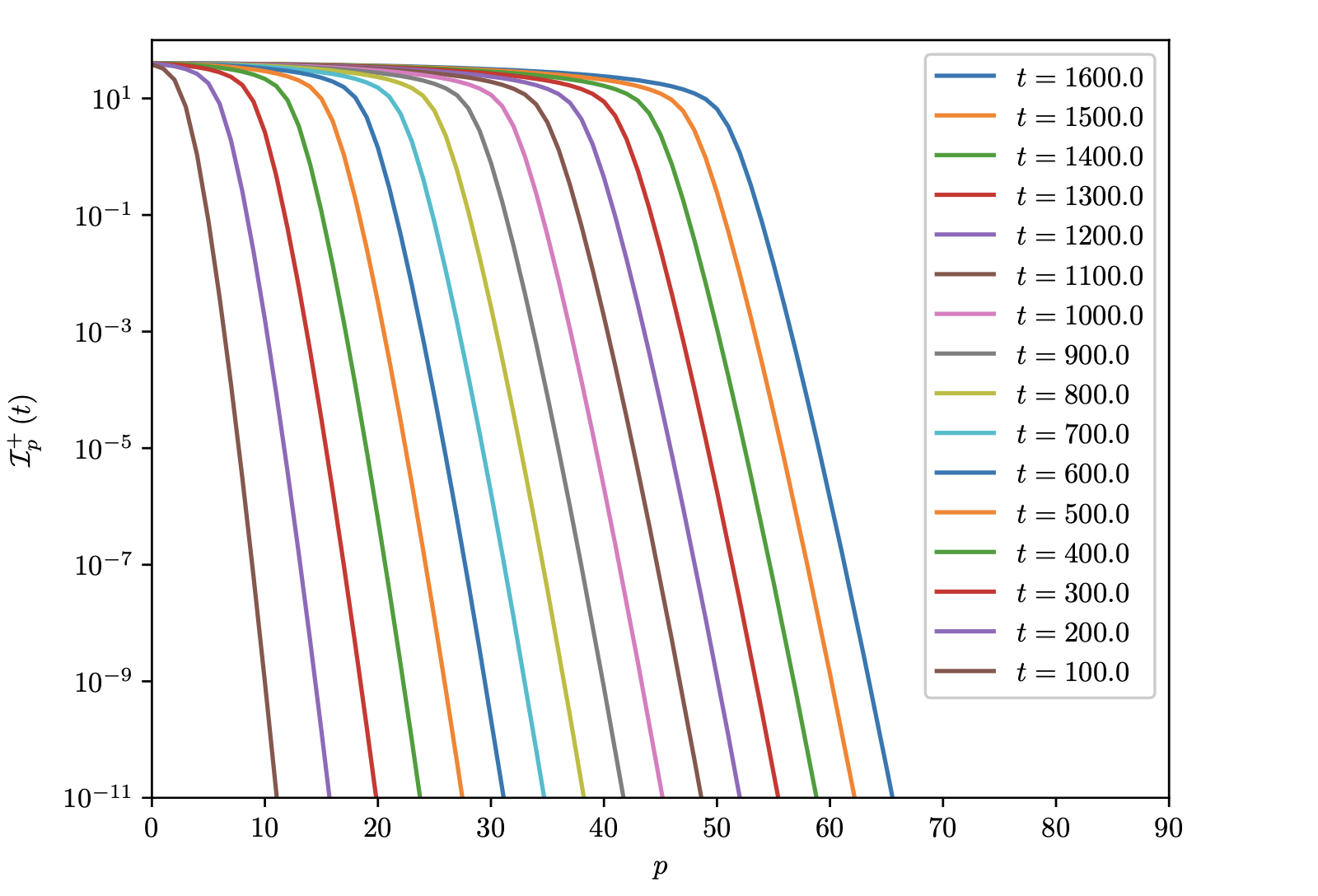}
\caption{\label{fig:Plots-of-eigenvalues_rho_plus}Plots of eigenvalues $\mathcal{I}_{p}^{+}\left(t\right)$ vs $p$ at a fixed $t$. The plots are presented for a number of values of $t$. It is seen that $\mathcal{I}_{p}^{+}\left(t\right)$ obey to some kind of Lieb-Robinson bound: there are first $Bt$ eigenvalues which are of order $\mathcal{I}_{1}^{+}\left(\infty\right)$. Subsequent eigenvalues decay at least exponentially fast: $\mathcal{I}_{Bt+\delta}^{+}\left(t\right)\protect\leq\mathcal{I}_{1}^{+}\left(\infty\right)\exp\left(-\gamma\delta\right)$. It is seen from the plot that $\gamma$ is asymptotically constant
at large $t$.}
\end{figure}

In Fig.~\ref{fig:Plots-of-eigenvalues_rho_plus} we present the plot of dependence of $\mathcal{I}_{p}^{+}\left(t\right)$ on $p$ at a fixed $t$, for a number of times $t$. The plotted dependency suggests that there is some kind of Lieb-Robinson bound: 
\begin{equation}
\mathcal{I}_{p}^{+}\left(t\right)\leq\mathcal{I}_{1}^{+}\left(\infty\right)\exp\left(-\gamma\theta\left(p-Bt\right)\right)\label{eq:interior_normal_modes_lieb_robinson_bounds}
\end{equation}
for some constant positive parameters $B$ and $\gamma$. Here $\theta$ is the Heaviside step function. 

Considering $m\left(t\right)=Bt+\delta$ light cone interior normal modes, the corresponding infidelity is bounded by: 
\begin{equation}
I\left( t\right)\lesssim\left[g\left\Vert \hat{V}_{s}\right\Vert \right]^{2}\mathcal{I}_{1}^{+}\left(\infty\right)e^{-\gamma\delta}
\label{eq:rel_guard_modes_convergence1}
\end{equation}

\section{Relevant subspace for local quench on a finite time interval\label{app:relevant_sub}}

Suppose we consider the evolution eq. (\ref{eq:initial_cond}-\ref{eq:interaction_pic_sch_eq}) on a finite time interval $\left[0,T\right]$. At the final time $T$ we find the light cone interior normal modes $\left|\widetilde{\kappa}_{1}\right\rangle \ldots|\widetilde{\kappa}_{m\left(T\right)}\rangle $ as described in Sec.~\ref{subsec:Lieb-Robinson-metric}. In fact, additional normal modes: the eigenstates $|\widetilde{\kappa}_{m\left(T\right)+p}\rangle $ for $p>0$ fall below the significance threshold, violating condition (\ref{eq:normal_mode_condition}). Owing to the light-cone consistency property (\ref{eq:light_cone_consistency}), these modes remain below the threshold for all times $t\in\left[0,T\right]$. Consequently, by choosing a sufficiently small threshold $a_{\rm cut}$, we can safely discard these modes from the Hamiltonian (\ref{eq:oqs_interaction_pic-1}) with only negligible distortion of computed observables $\propto O\left(\sqrt{a_{\rm cut}}\right)$ throughout the interval $\left[0,T\right]$.

To express $\hat{H}\left(t\right)$ in terms of the light-cone interior normal modes $\hat{\widetilde\kappa}_{p}^{\dagger}$, we perform the diagonalization of $\hat{\rho}_{+}\left(T\right)$:
\begin{equation}
\hat{\rho}_{+}\left(T\right)=\widetilde{U}\left[\begin{array}{ccc}
\mathcal{I}_{1}^{+} & 0 & \ddots\\
0 & \mathcal{I}_{2}^{+} & 0\\
\ddots & 0 & \ddots
\end{array}\right]\widetilde{U}^{\dagger},
\end{equation}
so that the mode operators are related by:
\[\hat{\widetilde\kappa}_{p}^{ \dagger}=\sum_{k=1}^{\infty}\widetilde{U}_{kp}\hat{a}_{k}^{\dagger}
\,,\quad\hat{a}_{p}^{\dagger}=\sum_{k=1}^{\infty}\widetilde{U}^{\dagger}_{pk}\hat{\widetilde\kappa}_{k}^{\dagger}=\sum_{k=1}^{\infty}\left[\widetilde\kappa_{k}^{*}\right]_{p}\hat{\widetilde\kappa}_{k}^{\dagger}
\]
where $[\widetilde\kappa_k^{*}]_p$ denotes the $p$th component of the eigenvector $\widetilde\kappa_k^{*}$. Substituting this into the spread equation (\ref{eq:spread_equation}) gives:

\begin{widetext}
\begin{equation}
\hat{a}_{0}^{\dagger}\left(t\right)=\sum_{p=0}^{\infty}\alpha_{p}\left(t\right)\sum_{k=1}^{\infty}\left[\widetilde\kappa_{k}^{*}\right]_{p}\hat{\widetilde\kappa}_{k}^{\dagger}=\sum_{k=1}^{\infty}\left\langle \widetilde\kappa_{k}\left|\alpha\left(t\right)\right.\right\rangle \hat{\widetilde\kappa}_{k}^{\dagger}
=\sum_{k=1}^{m\left(T\right)}\left\langle \widetilde\kappa_{k}\left|\alpha\left(t\right)\right.\right\rangle \hat{\widetilde\kappa}_{k}^{\dagger}+\sum_{k=m\left(T\right)+1}^{\infty}\left\langle \widetilde\kappa_{k}\left|\alpha\left(t\right)\right.\right\rangle \hat{\widetilde\kappa}_{k}^{\dagger}
\end{equation}
\end{widetext}
The second term is outside the light cone, and we neglect it. Substituting the result into the Hamiltonian~\eqref{eq:oqs_interaction_pic-1} gives:
\begin{equation}
\begin{split}
\hat{H}\left(t\right)\approx\hat{\widetilde{H}}\left(t\right) = \hat{H}_{s}+g\,\hat{V}_{s}^{\dagger}\sum_{k=1}^{m\left(T\right)}\left\langle \alpha\left(t\right)\left|\widetilde\kappa_{k}\right.\right\rangle \hat{\widetilde\kappa}_{k}+\\+g \, \hat{V}_{s}\sum_{k=1}^{m\left(T\right)}\left\langle \widetilde\kappa_{k}\left|\alpha\left(t\right)\right.\right\rangle \hat{\widetilde\kappa}_{k}^{\dagger}
\end{split}  
\label{eq:H_rel}
\end{equation}
When computing the local quench dynamics on the interval $\left[0,T\right]$, we solve the Schrödinger equation projected onto the light-cone interior: 
\begin{equation}
i\partial_{t}\left|\widetilde\Psi\left(t\right)\right\rangle=\hat{\widetilde{H}}\left(t\right)\left|\widetilde\Psi\left(t\right)\right\rangle.\label{eq:interaction_pic_sch_eq_lc_interior}
\end{equation}

\begin{figure}
\includegraphics[width=0.4\textwidth]{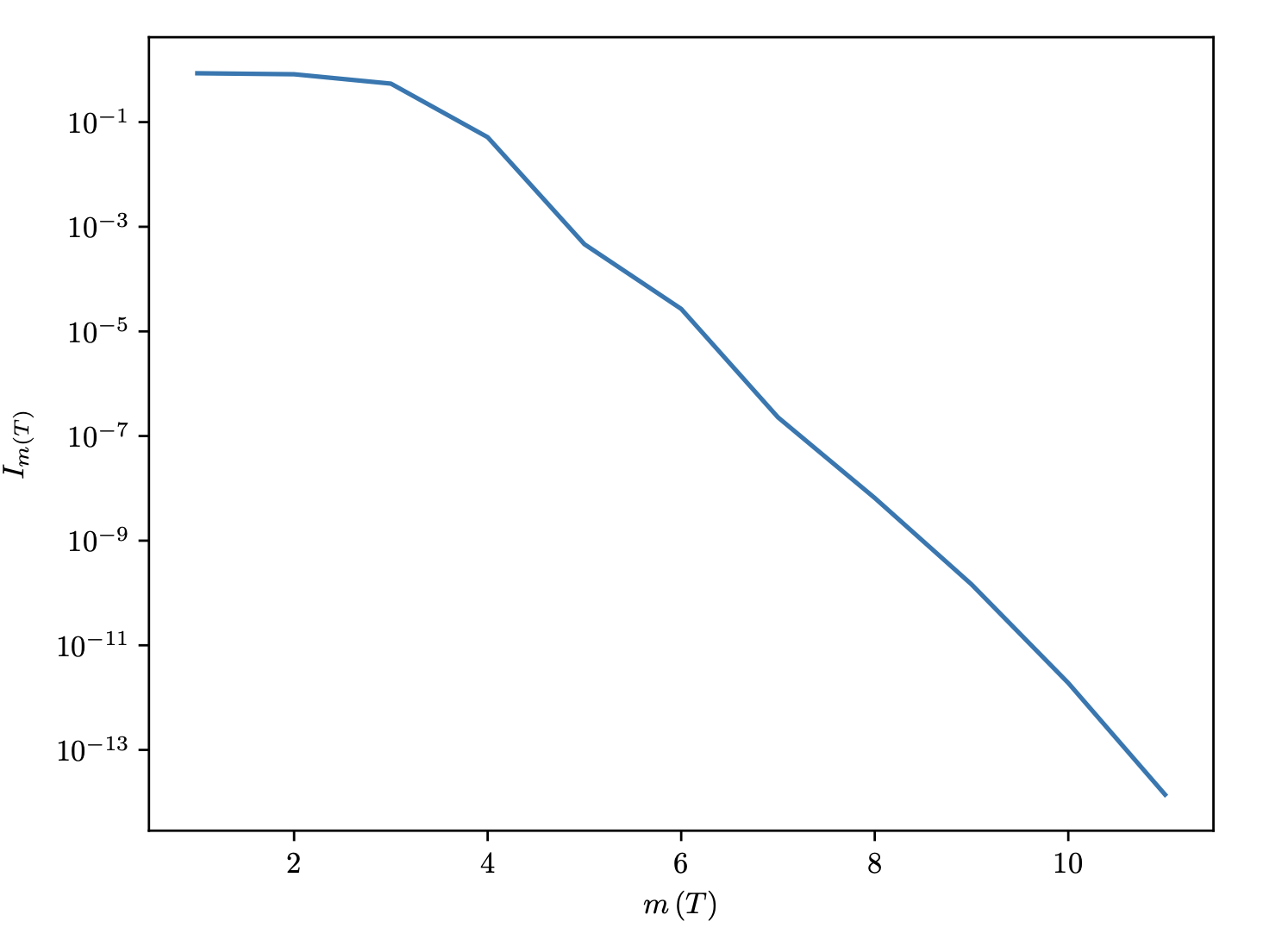}
\caption{\label{fig:Square-root-infidelity-between}Square-root infidelity between the wavefunction $|\widetilde\Psi\left(t\right)\rangle$ keeping only $m\left(T\right)$ light cone interior normal modes, and the wavefunction $\ket{\Psi\left(T\right)}$ containing all modes. The nonstationary Schrodinger equation was solved numerically in a Fock space which was truncated at a maximal number of environment's quanta $n_{cut}=4$ for $T=100$. The driven qubit $\hat{H}_{s}\left(t\right)=\hat{\sigma}_{+}\hat{\sigma}_{-}+\hat{\sigma}_{x}0.1\cos t$ was taken as an open system, $g=0.1$, $\hat{V}_{s}=\hat{\sigma}_{-}$, $\varepsilon_{j}\equiv1$, $h_{j}\equiv0.05$. The semiinfinite chain (\ref{eq:semiinf_chain}) was truncated at $30$ sites. Observe that $n_{cut}=4$ is not enough to avoid distortion of the computed wavefunction. Nevertheless, the computed ``truncated'' quantum dynamics is non-perturbative, and the infidelity estimate (\ref{eq:rel_guard_modes_convergence}) should hold provided $|\widetilde\Psi\left(t\right)\rangle$ and $\ket{\Psi\left(t\right)}$ are truncated in the same way.  }
\end{figure}

For numerical verification, we compute the square-root infidelity (\ref{eq:sqrt_fidelity}-\ref{eq:infidelity}), between the truncated state $\left|\widetilde\Psi\left(t\right)\right\rangle $ and the full state $\left|\Psi\left(t\right)\right\rangle $ for which all the normal modes $\hat{\widetilde\kappa}_{k}^{\dagger}$ are kept (the sums are not truncated at $m\left(T\right)$ in (\ref{eq:H_rel})). Fig.~\ref{fig:Square-root-infidelity-between} shows the infidelity $I_{m\left(T\right)}$ as a function of the number of retained modes $m\left(T\right)$.

\section{Finding minimal forward light cone basis \label{app:find_MFLCB}}

In this Appendix we present the explicit algorithm used to construct the minimal forward light-cone basis introduced in Sec.~\ref{subsec:Frame-of-a-minimal-lc}.

To obtain the minimal light cone, we apply a sequence of unitary rotations that delay the coupling of the least significant modes as much as possible. The construction proceeds backward in time.

At the final time $T$, all modes are coupled and their number equals $m_+(T)\equiv m(T)$. As we move backward in time, we solve the eigenvalue problem:
\begin{equation}
\rho_+(t)\ket{\bar{\kappa}_p(t)} =\mathcal{I}^+_p(t)\ket{\bar{\kappa}_p(t)}   
\end{equation}
with $\mathcal{I}_1^+(t)\ge\mathcal{I}_2^+\ge\ldots$ sorted in descending order. 
When the forward light-cone condition (\ref{eq:normal_mode_condition}) is violated for a particular eigenmode, $\mathcal{I}^+_k(t) - r_{\rm cut} \,\mathcal{I}^+_1(t) < 0$ ,
this mode is considered to decouple from the light cone. It is then added to the ordered sequence of incoming modes $\kappa_{k}^{\rm in} = \kappa_{m_+(t)}^{\rm in} =\bar{\kappa}_{k}(t)$, and the corresponding time $t_k^{\rm in}$ is recorded as its arrival time.

By recursively repeating this procedure while evolving backward in time, all coupled modes $\kappa^{\rm in}_k$ and their arrival times $t_k^{\rm in}$ are determined. The minimal forward light-cone basis is obtained from the sequence of eigenvector rotations generated during this process.

\section{Transformation to moving basis \label{app:trans_moving_basis}}
There are two competing processes as the system evolves in time. First, the new modes $\hat{\kappa}_{p}^{\rm in\dagger}$ enter the minimal forward light cone (Fig.\ref{fig:local_quench}) at specific time moments $t_{\rm in}\left(\kappa_{p}^{\rm in}\right)$. Second, the irreversibly decoupled modes $\hat{\kappa}_{p}^{\rm out\dagger}$ escape the interaction region at times $t_{\rm out}\left(\kappa_{p}^{out}\right)$. The system effectively evolves only with relevant modes, those that have coupled but have not yet irreversibly decoupled. Here we construct the time-dependent unitary transformation $W\left(t\right)$ to the moving frame of the relevant modes: 
\begin{equation}
\hat{\kappa}_{p}^{\rm mov \dagger}=\sum_{q=1}^{m(T)}W_{pq}(t)\hat{\kappa}_{q}^{\rm in \dagger} = \sum_{q=1}^{m(T)}W_{pq}(t)\sum_{l=1}^{m(T)}[U^{+ \rm min}]_{ql} \hat{\Tilde{\kappa}}_l^\dagger,
\end{equation}
$p,q = 1,...,m(T)$ and the inverse transformation: 
\begin{equation}
\hat{\kappa}_{q}^{\rm in \dagger}=\sum_{q=1}^{m(T)}W_{qp}^\dagger(t)\hat{\kappa}_{p}^{\rm mov \dagger},
\end{equation}
where $\hat{\kappa}_{p}^{\rm mov \dagger}$ is the sequence of independent degrees of freedom in the moving frame. For each $\hat{\kappa}_{p}^{\rm mov\dagger}$ there are two time moments: the arrival time $t^{\rm in}$ and the escape time $t^{\rm out}$.

Having the basis of coupling modes the moving frame is defined by the events of irreversible modes decoupling. The transformation $W$ is constant if there are no coupling or decoupling events occurring during the interval $t$. At a given time $t$, there are $m_{+}(t)$ modes inside the forward light cone (\ref{eq:m_plus}). To determine the moving basis and identify the irreversibly decoupled modes, we consider the matrix elements of $\hat{\rho}_{-}(t)$ (\ref{eq:future_OTOC}) in the subspace spanned by the coupling modes $\kappa_1^{\rm in},\,\ldots,\,\kappa_{m_+(t)}^{\rm in}$ (\ref{eq:rho_minus_proj}). At each time $t$ we solve the eigenvalue problem: 
\begin{equation}
\rho_-^{m_+(t)}(t)\ket{\bar{\kappa}_p(t)} = \mathcal{I}^-_p(t)\ket{\bar{\kappa}_p(t)}
\end{equation}
with $\mathcal{I}_{1}^{-}(t)\ge\mathcal{I}_{2}^{-}(t)\ge...\mathcal{I}_{m_+(t)}^{-}(t)$ sorted in descending order. 
There is a time moment when the  smallest eigenvalue becomes lower than the threshold: $\mathcal{I}_{m_+(t)}^{-}(t) - r_{\rm cut}\,\mathcal{I}_1^{-}(t) < 0$. This is the escape time for the first irreversibly decoupled mode $\hat{\kappa}_{1}^{\rm out\dagger}$. 
As soon as a mode decouples, the dimension of the matrix $\rho_-$ decreases by 1. At each decoupling time $t_k^{\rm out}$ it is projected onto the subspace of $r(t_k^{\rm out}) = m_{+}(t_k^{\rm out}) - m_{-}(t_k^{\rm out})$ modes, i.e., the relevant subspace. By recursively repeating this procedure, the stream of outgoing modes irreversibly decoupled from the system can be determined.

The moving basis is obtained recursively from the stream of incoming modes via time dependent transformation $W(t)$:
\begin{equation}
\left(\begin{array}{c}
\hat{\kappa}_1^{\rm mov\dagger} \\ \vdots \\ \hat{\kappa}_{m_+(t)}^{\rm mov\dagger}
\end{array}\right) =
\left(\begin{array}{c}
\hat{\kappa}_1^{\rm out\dagger} \\ \vdots \\ \hat{\kappa}_{m_-(t)}^{\rm out\dagger} \\ \hat{\kappa}_1^{\rm rel\dagger} \\ \vdots \\ \hat{\kappa}_{r(t)}^{\rm rel\dagger}
\end{array}\right) = W(t)
\left(\begin{array}{c}
\hat{\kappa}_1^{\rm in\dagger} \\ \vdots \\\hat{\kappa}_{m_+(t)}^{\rm in\dagger}
\end{array}\right)
\end{equation}

\begin{equation}
W_{pq}(t) = [\bar{\kappa}_p(t)]_q
\end{equation}
where $[\bar{\kappa}_p(t)]_q$ denotes the $q$-th component of eigenvector $\ket{\kappa^{\rm rel}_p(t)}$. Between two successive events $t_k^* < t < t_k^{\rm out}$, where $t_k^*$ is the time of the previous coupling or decoupling event, the transformation is constant: $W(t) = W(t_k^*)$.

At the escape time $t_k^{\rm out}$ the relevant rotation is implemented by:
\begin{equation}
W(t_k^{\rm out}) = \exp\big(-i (t_k^{\rm out}-t_k^*) \xi_k \big)
\end{equation}
where $\xi(t)$ is the generator of the transformation, which is piecewise constant $\xi_k$ on the interval $t_k^* < t < t_k^{\rm out}$. Thus, after several decoupling events the total transformation can be written as a product of successive rotations:
\begin{equation}
W(t) = W_k W_{k-1} \ldots W_1, \quad W_i = W(t_i^{\rm out}).
\end{equation}

\begin{widetext}
For clarity let us consider the time moment, then $k$ modes irreversibly decoupled:
\begin{equation}
\begin{split}
\left(\begin{array}{c}
\hat{\kappa}_1^{\rm out\dagger} \\ \vdots \\ \hat{\kappa}_k^{\rm out\dagger} \\ 
 \hat{\kappa}_1^{\rm rel\dagger} \\ \vdots \\ \hat{\kappa}_{r(t_k^{\rm out})}^{\rm rel\dagger}
\end{array}\right) =
W_k \ldots W_2 W_1 \left(\begin{array}{c}
\hat{\kappa}_1^{\rm in\dagger} \\ \vdots \\ \hat{\kappa}_{m_+(t_k^{\rm out})}^{\rm in\dagger}
\end{array}\right) = W_k \ldots \underbrace{\left(\begin{array}{cc}
1 & 0 \\
0 & W(t_2^{\rm out})
\end{array}\right)\left(\begin{array}{c}
\hat{\kappa}_1^{\rm out\dagger} \\ \hat{\kappa}^{\prime}{}_1^{\rm rel\dagger} \\ \vdots
\end{array}\right)}_{\left(\begin{array}{c}
\hat{\kappa}_1^{\rm out\dagger} \\ \hat{\kappa}_2^{\rm out\dagger} \\ \hat{\kappa}^{\prime\prime}{}_1^{\rm rel\dagger} \\ \vdots
\end{array}\right)}=\left(\begin{array}{cc}
\mathbbm{1}_{(k-1)\times(k-1)} & 0 \\
0 & W(t_k^{\rm out})
\end{array}\right)\left(\begin{array}{c}
\hat{\kappa}_1^{\rm out\dagger} \\ \vdots \\ \hat{\kappa}_{k-1}^{\rm out\dagger} \\ \hat{\kappa}^{\prime\prime\prime}{}_1^{\rm rel\dagger} \\ \vdots
\end{array}\right)
\end{split}
\end{equation}
\end{widetext}

So the effective Hamiltonian in the moving basis is:
\begin{equation}
\begin{split}
\hat{H}_{\rm eff}(t) = \hat{H}_s + g\,\hat{V}_{s}^{\dagger}\sum_{k=1}^{r(t)} \chi_{k}^{*}\left(t\right)\hat{\kappa}_{k}^{\rm rel}\,+\\+\,g\,\hat{V}_{s}\sum_{k=1}^{r(t)}\chi_{k}\left(t\right)\hat{\kappa}_{k}^{\rm rel}{}^\dagger - \sum_{k=1}^{r(t)}\sum_{l=1}^{r(t)}\xi_{kl}\left(t\right)\hat{\kappa}_{k}^{\rm rel}{}^\dagger\hat{\kappa}_{l}^{\rm rel}
\end{split}
\end{equation}
where 
\begin{equation}
\chi_{k}(t) = \sum_{p=0}^{\infty}\alpha_p W^*_{pk}(t) = \bra{\kappa_k^{\rm rel}}\alpha\rangle   
\end{equation} are the time-dependent coupling amplitudes and
\begin{equation}
\xi_{kl}(t^{\rm out}_p) = \left[i \ln{W(t^{\rm out}_p)}/(t^{\rm out}_p - t^*)\right]_{kl}\,\,.    
\end{equation}

The infidelity bound for moving basis constructed analogous with infidelity of forward light cone.

The Hamiltonian for the discarded mode:
\begin{equation}
\Delta \hat{H} = \hat{H} - \hat{H}_{\rm eff} = g\,\hat{V}_{s}^{\dagger} \chi^{*}\left(t\right)\hat{\kappa}\,
+
\,g\,\hat{V}_{s}\chi\left(t\right)\hat{\kappa}^\dagger
\end{equation}
since $\xi$ acts only in the relevant subspace. 

Thus all the calculation remind the Appendix ~\ref{app:just_Lieb_Rob}:
\begin{equation}
1 -  \left| \langle\Psi_{\rm eff}(t)|\Psi(t)\rangle\right| \lesssim\left[g\left\Vert \hat{V}_{s}\right\Vert \right]^{2}\mathcal{I}_{1}^{+}\left(\infty\right)e^{-\gamma\delta}
\label{eq:INFID}
\end{equation}
where $|\Psi_{\rm eff}(t)\rangle$ evolves with $H_{\rm eff}(t)$.

\begin{figure}[bp]
\centering
\includegraphics[width=0.4\textwidth]{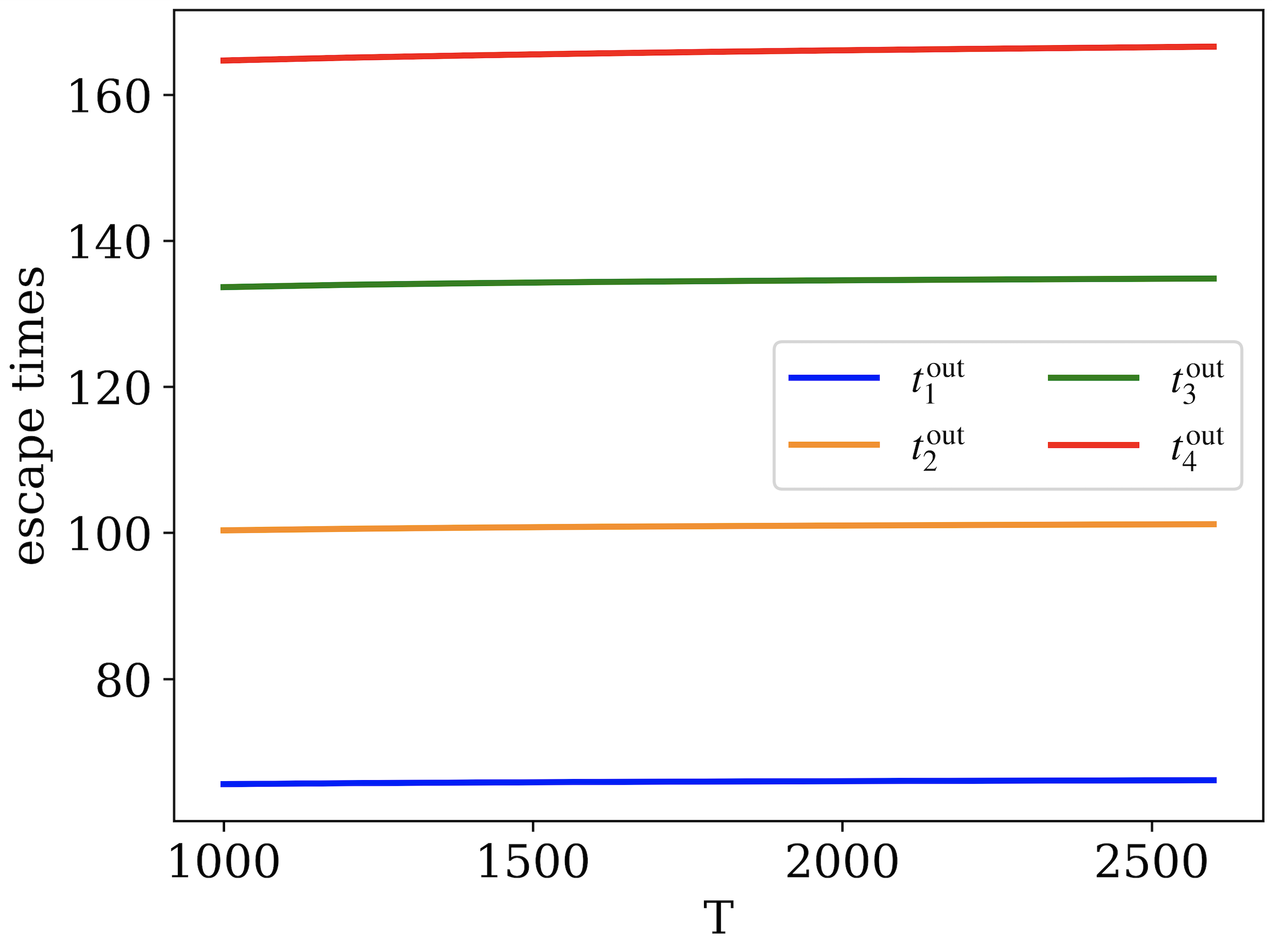}
\caption{Escape time $t_k^{\rm out}$ as a function of the total evolution time $T$.}
\label{fig:t_out_vs_T}
\end{figure}

\begin{figure*}
\centering
\includegraphics[width=0.52\textwidth]{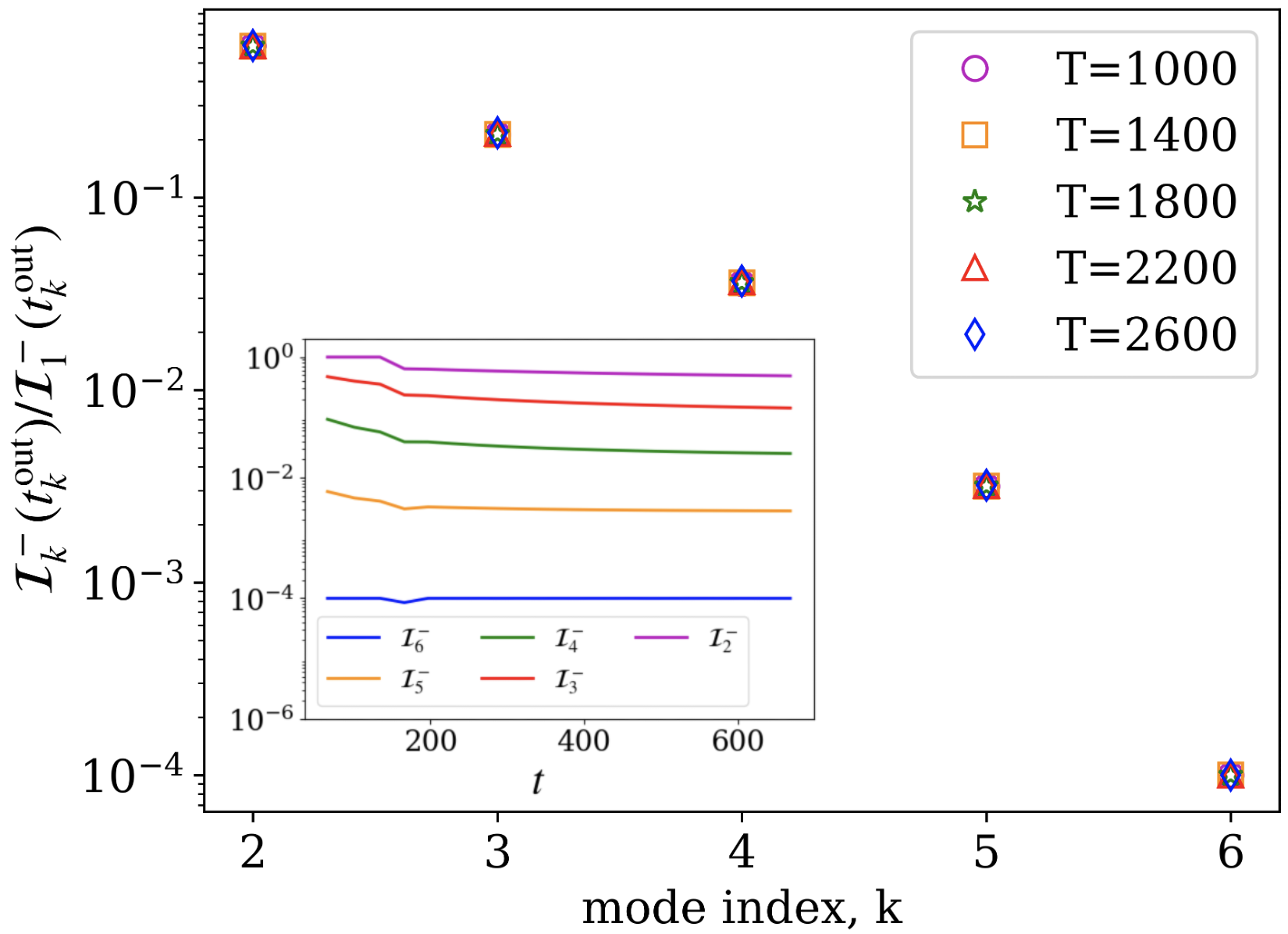}
\hfill
\includegraphics[width=0.45\textwidth]{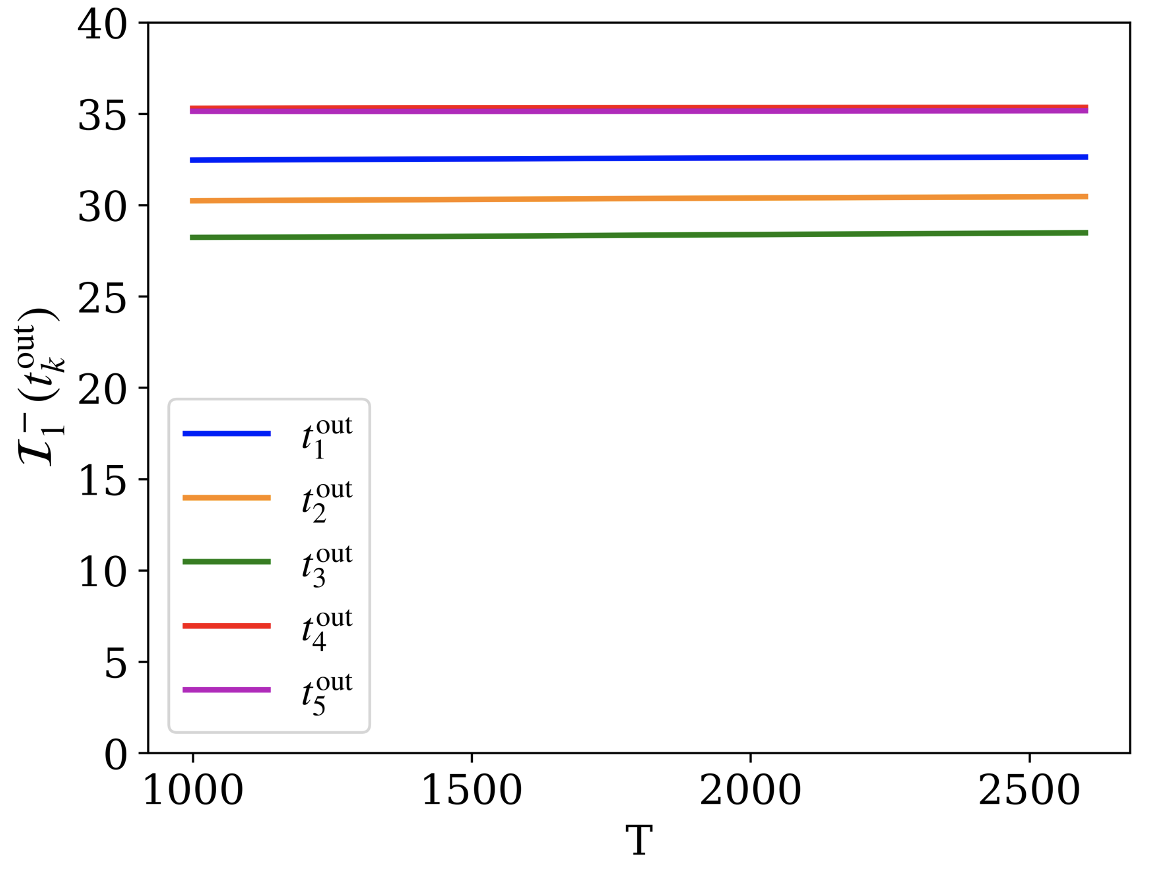}
\caption{The left figure shows spectrum of $[\hat{\rho}_-^{m_+(t)}(t)]_{pq} = \langle\kappa_p^{\rm in}|\hat{\rho}_-(t)|\kappa_p^{\rm in}\rangle$ at a fixed decoupling event $t_k^{\rm out}$ for different total evolution times $T$, 
normalized by the largest eigenvalue. The inset plot shows the time evolution of the normalized eigenvalues $\mathcal{I}_j^-(t_k^{\rm out})/\mathcal{I}_1^-(t_k^{\rm out})$ evaluated at the decoupling times $t_k^{\rm out}$. The right figure demonstrates dependence of the largest eigenvalue $\mathcal{I}_1^-(t_k^{\rm out})$ on the total evolution time $T$ for several decoupling events indexed by $k$.}\label{fig:TT}
\end{figure*}

Fig.~\ref{fig:t_out_vs_T} shows the escape time $t_k^{\rm out}$ as a function of the total evolution time $T$. Within the considered range, the escape times are nearly independent of $T$. Doubling $T$ changes $t_k^{\rm out}$ by less than $1\%$. This demonstrates that the escape times are determined by the intrinsic dynamics of the system and are not sensitive to the truncation of the evolution time $T$. One can see (Fig.~\ref{fig:TT}, left plot) that the spectral structure of $[\hat{\rho}_-^{m_+(t)}(t)]_{pq} = \langle\kappa_p^{\rm in}|\hat{\rho}_-(t)|\kappa_p^{\rm in}\rangle$ remains stable over time and the lowest mode remains at the level $\sim r_{\rm cut}$ and independent of $T$, consistent with the decoupling criterion (\ref{eq:backwardLC}). The largest eigenvakue of $[\hat{\rho}_-^{m_+(t)}(t)]_{pq}$: $\mathcal{I}_1^-(t_k^{\rm out})$ remains constant as $T$ increases, indicating the saturation with respect to the total evolution time (Fig.~\ref{fig:TT}, right plot).  This demonstrates that no accumulation of contributions from late times occurs and provides the evidence for the absence of a finite breakdown timescale $T_{\rm break}$.

\section{Error bound for commutators of emerging integrals of motion \label{app:err_bound}}

We estimate the commutator $\left[\hat{H}_{\rm int}(t), \hat{P}_{\kappa^{\rm out};\, n_{\kappa^{\rm out}}} \left(t^{\rm out}\right)\right]$ at the time $t^{\rm out}$ when the mode $\kappa^{\rm out}$ becomes irreversibly decoupled.

At $t=t^{\rm out}$ the interaction Hamiltonian reads
\begin{equation}
\hat{H}_{\rm int}\left(t^{\rm out}\right) =  g\,\hat{V}_{s}^{\dagger} \hat{a}_0(t^{\rm out}) + g\,\hat{V}_{s} \hat{a}^{\dagger}_0(t^{\rm out})
\end{equation}

We decompose the the system-environment coupling into the component along the outgoing mode $\kappa^{\rm out}$ and an orthogonal part:
\begin{equation}
\hat{a}_{0}^{\dagger}\left(t^{\rm out}\right)=\hat{a}_{\perp}^{\dagger}\left(t^{\rm out}\right)+
\left\langle \kappa^{\rm out}\left|\right.\alpha\left(t^{\rm out}\right)\right\rangle \hat{\kappa}^{\rm out}\,.
\end{equation}

\begin{figure}
\includegraphics[width=0.4\textwidth]{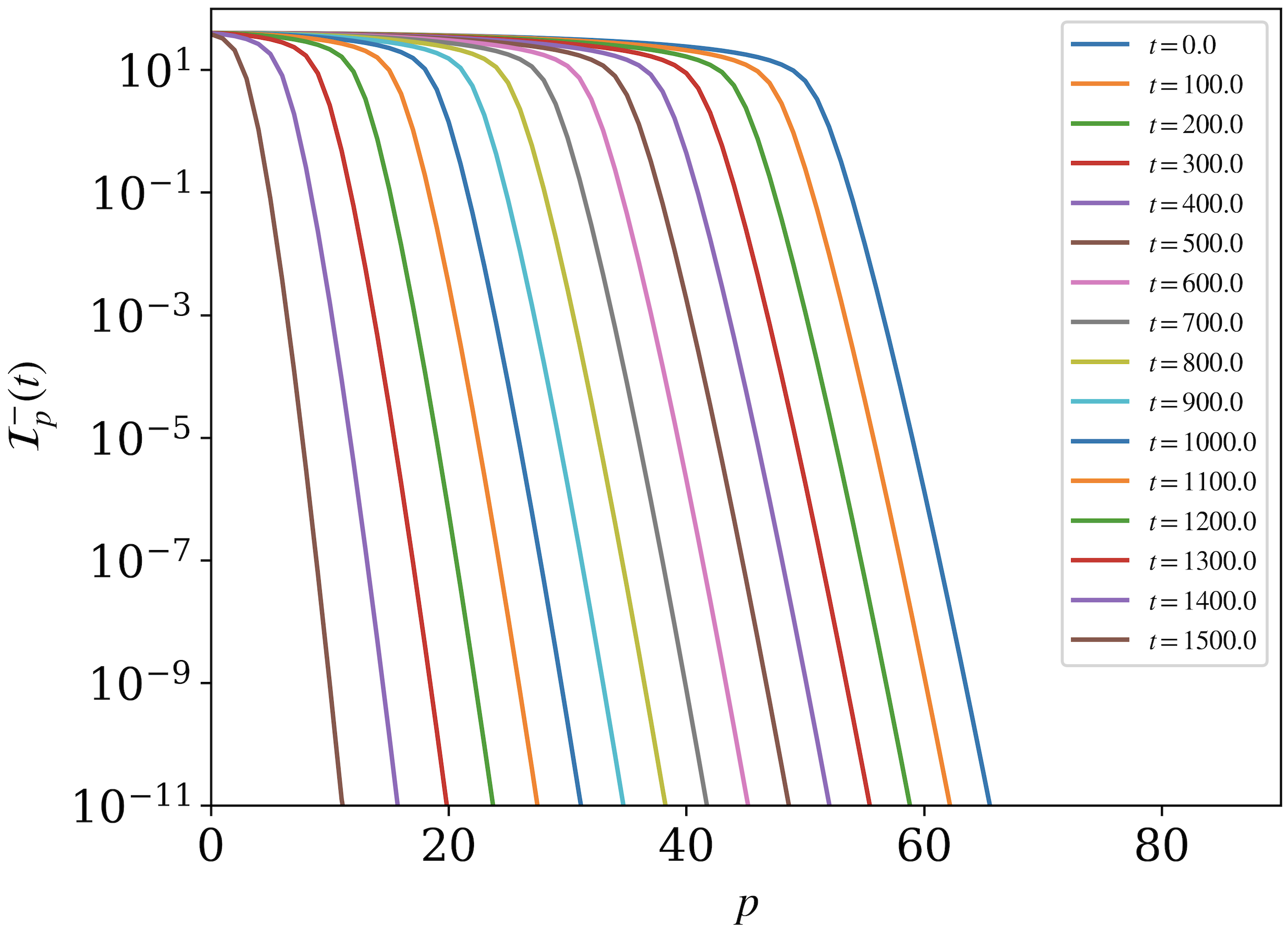}
\caption{Plots of eigenvalues $\mathcal{I}_p^{-}\left(t\right)$ of $\hat{\rho}_{-}\left(t\right)$ vs $p$ at a fixed $t$ are presented for a number of values of $t$, demonstrates some kind of Lieb–Robinson structure for the backward light-cone.
Similar to the Fig.\ref{fig:Plots-of-eigenvalues_rho_plus}.}
\label{fig:eigenvalues_rho_min}
\end{figure}

Since $\hat{a}_\perp$ acts in the subspace orthogonal to $\kappa^{\rm out}$, it commutes with the projector $\hat{P}_{\kappa^{\rm out};\, n_{\kappa^{\rm out}}}$. Therefore, only the $\hat{\kappa}^{\rm out}$ component contributes to the commutator.

\begin{widetext}
Using the definition (\ref{eq:eigenspace}) we obtain:
\begin{equation}
\begin{gathered}
\left[\hat{H}_{\rm int}, \hat{P}_{{\kappa^{\rm out}};\, n_{\kappa^{\rm out}}}\right] = 
\left[\hat{H}_{\rm int},\ket{n^{\rm out}}\bra{n^{\rm out}}\right] = \\
= g \hat{V}^\dagger_s\,\langle \kappa^{\rm out}\left|\alpha\left(t^{\rm out}\right)\right\rangle\left( \sqrt{n^{\rm out}+1}\ket{n^{\rm out}+1}\bra{n^{\rm out}} - \sqrt{n^{\rm out}}\ket{n^{\rm out}}\bra{n^{\rm out}-1}\right) +
\\
+g \hat{V}_s\,\left\langle \alpha\left|\right.\kappa^{\rm out}\left(t^{\rm out}\right)\right\rangle\left( \sqrt{n^{\rm out}}\ket{n^{\rm out}-1}\bra{n^{\rm out}} - \sqrt{n^{\rm out}+1}\ket{n^{\rm out}}\bra{n^{\rm out}+1}\right)
\label{eq:commutator_explicit}
\end{gathered}
\end{equation}
\end{widetext}

Taking the operator norm yields the bound:
\begin{equation}
\left|\left[\hat{H}_{\rm int}, \hat{P}_{{\kappa^{\rm out}};\, n_{\kappa^{\rm out}}}\right]\right|  \lesssim C \left|g\right|\left\Vert \hat{V}_{s}\right\Vert \left|\left\langle \kappa^{\rm out}\left|\right.\alpha\left(t^{\rm out}\right)\right\rangle\right|
\end{equation}
where $C$ is a positive constant controlled by the maximal occupation 
number of the outgoing mode $\kappa^{\rm out}$. This occupation number of the order of number of quanta emitted into that mode, approximately is $\propto \rm intensity \times bandwidth$.

From the backward light-cone condition (\ref{eq:backwardLC}) and (\ref{eq:a_cut}), (\ref{eq:rel_guard_modes_convergence}) we have: 
\begin{equation}
\int_{t^{\rm out}}^T d\tau \left| \langle \kappa^{\rm out}|\alpha(\tau)\rangle \right|^2 < \mathcal{I}_1^{-}(\infty)\, r_{\rm cut}.
\end{equation}

The spectrum of $\hat{\rho}_{-}\left(t\right)=\int\limits_{t}^{T}d\tau\left|\alpha\left(\tau\right)\right\rangle \left\langle \alpha\left(\tau\right)\right|$ follows from those the spectrum $\hat{\rho}_{+}\left(t\right)=\int\limits_{0}^{t}d\tau\left|\alpha\left(\tau\right)\right\rangle \left\langle \alpha\left(\tau\right)\right|$ by time reflection $\rho_-(t) = \rho_+(T-t)$ and exhibits a Lieb–Robinson type bound: the number of significant eigenvalues scales linearly with $T-t$, while higher modes are exponentially suppressed. For mode indices beyond the effective light-cone: 
$m = B (T - t) + \delta$ , $\delta \ge 0 $,
the corresponding intensity satisfies:
\[
\mathcal I_m^{-}(t)
\le
\mathcal I_1^{-}(\infty)\, e^{-\gamma \delta}.
\]
with $I_1^{-}(\infty)$ saturation level similar to $I_1^{+}(\infty)$ (Fig.~\ref{fig:Plot-of-the_I1_plus_t}). We therefore identify $r_{\rm cut} = e^{-\gamma \delta}$.

Thus, the amplitude 
$\langle \kappa^{\rm out}|\alpha(t)\rangle$ can vary only on the characteristic timescale:

\begin{equation}
t^* \sim \frac{\delta}{B} = \frac{-\ln r_{\rm cut}}{\gamma B}.
\end{equation}

Combining the integral bound with this timescale yields:

\begin{equation}
\left| \langle \kappa^{\rm out}|\alpha(t^{\rm out})\rangle \right| \lesssim {\rm const} \sqrt{ \frac{r_{\rm cut}}{-\ln r_{\rm cut}} } \le {\rm const} \sqrt{r_{\rm cut}}
\end{equation}

The final estimation is:
\begin{equation}
\left\Vert\left[\hat{H}_{\rm int}(\tau), \hat{P}_{{\kappa^{\rm out}};\, n_{\kappa^{\rm out}}}\right] \right\Vert \lesssim {\rm const} \left| g \right|\left\Vert \hat{V}_{s}\right\Vert \sqrt{r_{\rm cut}}
\end{equation}
for $\tau>t_\kappa^{\rm out}$
This implies $\left[\hat{U}(t_2, t_1),\, \hat{P}_{\kappa_1^{\rm out};\, n_1^{\rm out}}\right] = O(\sqrt{r_{\rm cut}})$, where $t_2$ and $t_1$ is fixed and independent of $T$ (Fig.~\ref{fig:TT}).

Then we can estimate the bound of decoherence functional.
For histories $\boldsymbol{\alpha} = (a_1, a_2)$ and $\boldsymbol{\beta} = (a_1', a_2')$ the decoherence functional reads:
\begin{equation}
\begin{gathered}
\mathcal{D}(\boldsymbol{\alpha},\boldsymbol{\beta}) = \langle\Psi(0)|  \hat{U}^\dagger(t_1)\,\hat{P}_{1,a_1'}\,\hat{U}^\dagger(t_2,t_1)\,\\
\hat{P}_{2,a_2'}\,
\hat{P}_{2,a_2}\,\hat{U}(t_2,t_1)\,\hat{P}_{1,a_1}\,\hat{U}(t_1)|\Psi(0) \rangle
\label{eq:D_two_jump}
\end{gathered} 
\end{equation}

Using the notation $\hat{\varepsilon}_{a_1} = [\hat{U}(t_2,t_1),\hat{P}_{1,a_1}]$
and substituting into \ref{eq:D_two_jump} (with $\delta_{a_2,a_2'}=1$ from (\ref{eq:mutual_ex+ex})):
\begin{equation}
\begin{gathered}
\mathcal{D}(\boldsymbol{\alpha},\boldsymbol{\beta}) = \delta_{a_1,a_1'} 
P(\boldsymbol{\alpha}) + \\
+ \langle \Psi(0) | \hat{U}^\dagger(t_1)\,\hat{U}^\dagger(t_2,t_1)\,
\hat{P}_{1,a_1'}\,\hat{P}_{2,a_2}\,\hat{\varepsilon}_{a_1}\,\hat{U}(t_1)|\Psi(0) \rangle +\\
+ \langle \Psi(0) |\hat{U}^\dagger(t_1)\,\,\hat{\varepsilon}_{a_1'}^\dagger\,
\hat{P}_{2,a_2}\,\hat{P}_{1,a_1}\,\hat{U}(t_2,t_1)\,\,\hat{U}(t_1)|\Psi(0) \rangle+\\
+ \langle \Psi(0) |\hat{U}^\dagger(t_1)\,\hat{\varepsilon}_{a_1'}^\dagger\,
\hat{P}_{2,a_2}\,\hat{\varepsilon}_{a_1}\,\hat{U}(t_1)|\Psi(0) \rangle.
\label{eq:expansion}
\end{gathered}
\end{equation}

This yields the following bound for decoherence functional:
\begin{equation}
\mathcal{D}(\boldsymbol{\alpha},\boldsymbol{\beta})(t)\lesssim \left\Vert\left[\hat{H}_{\rm int}, \hat{P}_{{\kappa^{\rm out}};\, n_{\kappa^{\rm out}}}\right] \right\Vert = O(\sqrt{r_{\rm cut}})\,, \rm for\,\,\boldsymbol{\alpha}\neq \boldsymbol{\beta}
\end{equation}

One can obtain the estimation for decoherence functional using infidelity bound. We denote the normalized history state:
\begin{equation}
\ket{\Psi_{\boldsymbol{\alpha}}} = \ket{\Psi_{\perp \boldsymbol{\alpha}}} + \ket{\Delta\Psi_{ \boldsymbol{\alpha}}}
\end{equation}
where $\ket{\Psi_{\perp \boldsymbol{\alpha}}}$ is ideal decoherent history and $\ket{\Psi_{ \boldsymbol{\alpha}}}$ is defined in eq.(\ref{eq:hist_modes}). With $\Vert \Psi_{\boldsymbol{\alpha}}\Vert$ = $\Vert \Psi_{\perp \boldsymbol{\alpha}}\Vert$ = 1.

Using the bound (\ref{eq:INFID}) the infidelity between the actual and the ideal histories is:
\begin{equation}
I = 1 -  \left|\langle\Psi_{\perp \boldsymbol{\alpha}}(t)|\Psi_{\boldsymbol{\alpha}}(t)\rangle\right| \lesssim\left[g\left\Vert \hat{V}_{s}\right\Vert \right]^{2}\mathcal{I}_{1}^{+}\left(\infty\right)\, r_{\rm cut}
\end{equation}

Thus, $|\langle \Psi_{\perp \boldsymbol{\alpha}} | \Delta\Psi_{\boldsymbol{\alpha}} \rangle| = O(I)$ and:
\begin{equation}
1 = \Vert\Psi_{\perp\boldsymbol{\alpha}} \Vert^2 + 2 \Re\langle\Psi_{\perp\boldsymbol{\alpha}}|\Delta\Psi_{\boldsymbol{\alpha}}\rangle + \Vert\Delta\Psi_{\boldsymbol{\alpha}} \Vert^2 
\end{equation}
which implies $\Vert\Delta\Psi_{\boldsymbol{\alpha}}\Vert = O(\sqrt{I})$.

For decoherence functional (\ref{eq:DO_modes}) we have: 
\begin{equation}
\begin{gathered}
\mathcal{D}(\boldsymbol{\alpha},\boldsymbol{\beta})(t)=\bra{\Psi_{\boldsymbol{\beta}}(t)}\left.\Psi_{\boldsymbol{\alpha}}(t)\right\rangle = \\
=
\bra{\Psi_{\perp \boldsymbol{\beta}}(t)}\left.\Psi_{\perp \boldsymbol{\alpha}}(t)\right\rangle + \bra{\Delta\Psi_{\boldsymbol{\beta}}(t)}\left.\Psi_{\perp\boldsymbol{\alpha}}(t)\right\rangle + \\
+ \bra{\Psi_{\perp\boldsymbol{\beta}}(t)}\left.\Delta\Psi_{\boldsymbol{\alpha}}(t)\right\rangle + \bra{\Delta\Psi_{\boldsymbol{\beta}}(t)}\left.\Delta\Psi_{\boldsymbol{\alpha}}(t)\right\rangle
\end{gathered}
\end{equation}

By construction $\bra{\Psi_{\perp \boldsymbol{\beta}}(t)}\left.\Psi_{\perp \boldsymbol{\alpha}}(t)\right\rangle = 0$ for $\boldsymbol{\alpha}\neq\boldsymbol{\beta}$, then the decoherence functional:
\begin{equation}
\begin{gathered}
\mathcal{D}(\boldsymbol{\alpha},\boldsymbol{\beta})(t) \le \Vert \Delta\Psi_{\boldsymbol{\alpha}}(t) \Vert + \Vert \Delta\Psi_{\boldsymbol{\beta}}(t) \Vert +\\
+ \Vert \Delta\Psi_{\boldsymbol{\alpha}}(t) \Vert \Vert \Delta\Psi_{\boldsymbol{\beta}}(t) \Vert = O(\sqrt{I}) = O(\sqrt{r_{\rm cut}})\,\,\rm for \,\, \boldsymbol{\alpha}\neq\boldsymbol{\beta}
\end{gathered}
\end{equation}

\end{document}